# Collective motion of cells: from experiments to models

Elöd Méhes[a] and Tamás Vicsek*[a,b]

## Abstract

Swarming or collective motion of living entities is one of the most common and spectacular manifestations of living systems having been extensively studied in recent years. A number of general principles have been established. The interactions at the level of cells are quite different from those among individual animals therefore the study of collective motion of cells is likely to reveal some specific important features which we plan to overview in this paper. In addition to presenting the most appealing results from the quickly growing related literature we also deliver a critical discussion of the emerging picture and summarize our present understanding of collective motion at the cellular level. Collective motion of cells plays an essential role in a number of experimental and real-life situations. In most cases the coordinated motion is a helpful aspect of the given phenomenon and results in making a related process more efficient (e.g., embryogenesis or wound healing), while in the case of tumor cell invasion it appears to speed up the progression of the disease. In these mechanisms cells both have to be motile and adhere to one another, the adherence feature being the most specific to this sort of collective behavior. One of the central aims of this review is both presenting the related experimental observations and treating them in the light of a few basic computational models so as to make an interpretation of the phenomena at a quantitative level as well.

## Introduction

In this introductory section and in the section titled "Need for quantitative description" we provide the basic definitions of the notions used throughout the manuscript. Many of these were originally introduced for the level of organisms. Flocks of birds, herds or fish schools are perhaps the best known examples for large groups exhibiting fascinating patterns of motion by coordinating their motion in various ways (for extensive review see: [1]). Interestingly enough, some of the approaches developed for organisms can also be applied to the description of collective motion of cells as well. Although in some cases we make use of the terminology commonly accepted by the community studying the migration of cells, yet due to our focus on the quantitative interpretation of the related processes, we find that providing an introduction to the quantities and the basic models used throughout this review should be useful for the reader.

[a] Department of Biological Physics, Eötvös University, Budapest, Hungary. Fax: 36 13722757; Tel: 36 13722795; E-mail: emehes@angel.elte.hu

[b] MTA-ELTE Statistical and Biological Physics Research Group of the Hungarian Academy of Sciences, Budapest, Hungary. Fax: 36 13722757; Tel: 36 13722795; E-mail: vicsek@hal.elte.hu

**Defining collective cell motion**

Collective motion is a form of collective behavior: individual units (cells) interact in simple (attraction/repulsion) or complex way (through combination of simple interactions). The main feature of collective behavior is that the individual cell's action is dominated by the influence of other cells so that it behaves very differently from how it would behave if it was alone. The pattern of behavior is determined by the collective effects due to the other cells of the system.

For purposes of this review we emphasize two major characteristics of collective cell motion (migration). 1) Cells are physically and functionally connected with each other and connection is maintained during collective motion; 2) These multicellular structures exhibit polarity and the supracellular organization of individual cytoskeletal structures generates traction and protrusion forces for migration.

Although it is tempting to see the migration of loosely associated groups, e.g. germ cells, as a collective, however they are essentially solitary cells following the same (e.g.chemotactic) cues and tracks while occasionally contacting each other. Therefore we will not consider the migration of these groups as real collective migration because there is an apparent lack of collective effects.

Collective cell motion can occur in the form of 2-dimensional migration on a tissue surface or as 3-dimensional migration of a multicellular group (also termed: cohort) through a tissue scaffold. In the following we will provide a naturally incomplete list of selected examples for the observed subtypes collective migration from among higher eukaryotes in the context where they are experimentally studied: in embryonic development, wound-healing, vascular and tracheal network formation and *in vitro* conditions. Next we will collect, where available, some computational models trying to reproduce and explain the experimentally observed phenomena. Again, their list is rather exemplary and incomplete. Additionally, we will guide readers through the field of pattern formation by segregation of collectively moving cells where numerous computational models have been developed and tested.

**Main types of collective cell motion**

Collective cell migration in two dimensions is perhaps best exemplified by the sheet migration of fish keratocytes (skin cells) isolated from scales[2], the density-dependent sheet migration of isolated human endothelial cells (lining the inner surface of blood vessels) in culture during wound-healing[3] and the streaming behavior of endothelial cells in dense, confluent monolayers.[4]

There are several experimentally observed forms of 3-dimensional collective migration, mostly in morphogenic events. During the gastrulation process in the zebrafish embryo leading to the formation of the mesendodermal (germ line) layer, cells exhibit concerted 3D laminar migration.[5] The primordium of the lateral line organ migrates as one cohesive group with front and rear polarity in a later stage of zebrafish embryonic development giving rise to the chain of mechanosensory organs.[6] Similarly, polarized multicellular strands move collectively during branching morphogenesis of the mammary gland or the fruit fly's tracheal network.[7] The branching morphogenesis of the vascular network of a wide range of species from birds to mammals is also a known example of collective migration of polarized multicellular strands that are forming a tubular network.[8]

A somewhat special form of 3-dimensional collective migration is the migration of the completely isolated group of border cells towards the oocyte through the tissue stroma made up of nurse cells in the developing egg chamber of the fruit fly.[9]

During collective invasion observed in several human cancer types, such as epithelial cancers and melanomas, detached cell groups with front/rear polarity can migrate across tissues after tissue remodeling by the secretion of metalloprotease enzymes, cleaving the extracellular matrix. In some cancer types, the groups can switch among states ranging from collective migration through partial to complete individual migration in processes termed epithelial-mesenchymal transition (EMT) or mesenchymal-epithelial transition (MET). Their motion is reminiscent of morphogenic events but in a rather dysregulated way with the mechanisms yet to be understood making collective cancer invasion a field of great medical importance but more difficult to study compared to morphogenesis. Excellent reviews have been published on various aspects of collective cancer migration.[10,11,12,13]

Another interesting domain of life where collective motion is observed and modeled is the world of bacteria. Autonomously moving bacteria rely on motility organelles such as flagella or cilia making their motion very different from the collective motion of adherent tissue cells from higher animals that this Review is focusing on. Although the collective motion of bacteria falls outside the scope of this Review, a very detailed recent review on collective motion emerging at various organizational levels of life offers a good opportunity for comparison.[1]

**Need for quantitative description**

So far the collective motion of cells was mainly investigated by experimentalists and the corresponding reviews were concentrating on the phenomenological aspects of the related processes. In the second part of this Review we bring into the picture a number of computational models that can be successfully used to quantitatively interpret the observations. The quantitative treatment can be useful from the point of the understanding of the basics, but it has potential relevance to designing further experiments or even treatments in the case of cancer therapies.

Throughout this Review we use the terms collective motion, swarming, flocking or cohort migration as synonyms of coherent or ordered motion of units. In various models, collective motion is an emergent phenomenon arising from disordered, random motion through a transition as a function of relevant parameters of the system. Units of a system where collective motion emerges are i) rather similar, ii) moving with similar velocities and capable of changing their direction, iii) interacting with each other causing effective alignment of motion and iv) subject to perturbations from their environment.

The extent to which the motion of a population is collective is best indicated by a suitably chosen *order parameter*. The order parameter in this case is $\varphi$, the moving units' averaged velocity normalized between 0 and 1 as:

$$\varphi = \frac{1}{N v_0} \left| \sum_{i=1}^{N} \vec{v}_i \right|, \quad (Eq.1)$$

where N is the number of units, $v_0$ is the average absolute velocity of the units and $\vec{v}_i$ is the actual velocity of unit *i*.

If motion is disordered, the order parameter will be close to 0, whereas in case of ordered motion it will be close to 1. In experimental work, the actual velocity of individual cells can be measured using various methods ranging from manual tracking to automatic tracking based on e.g. object recognition or particle image velocimetry (PIV).

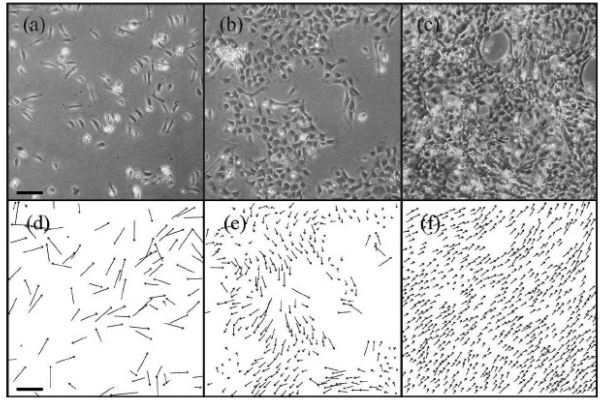

**Fig. 1** Sheet migration of epithelial cells in vitro.
Phase contrast images showing the collective behavior of primary goldfish keratocytes for three different densities. The normalized density, $\langle \rho \rangle$, is defined as $\langle \rho \rangle = \rho_{observed} / \rho_{max}$, where $\rho_{max}$ is the maximal observed density: 25 cells/100 × 100 micron area. (a) $\langle \rho \rangle = 0.072$ (b) $\langle \rho \rangle = 0.212$ and (c) $\langle \rho \rangle = 0.588$. Scale bar indicates 200 μm. As cell density increases, cell motility undergoes transition to collective ordering. The speed of coherently moving cells is smaller than that of solitary cells. (d)-(f) depict the corresponding velocities of the cells. From Szabó et al., (2006) with permission of Phys Rev E.

## Observations

### Collective cell motion in vitro

#### Sheet migration

This type of motion is primarily observed in the form of *in vitro* experiments in which the cells move on a plastic or glass surface, typically coated with a layer of proteins facilitating the motion (e.g., extracellular matrix proteins).

As a very characteristic form of 2-dimensional collective motion, the collective migration of keratocytes isolated from goldfish scales was studied by Szabó et al.[2] Based on the experimentally observed phenomenon of density-dependent ordering transition from individual random migration to ordered collective migration they determined this phase transition event as a function of cell density (**Fig. 1**). This was found to be continuous (second order) transition occurring as cell density exceeded a relatively well-defined critical value (also see Reference video 1).

Endothelial and epithelial cells are other cell types that have been used for studying in vitro 2-dimensional collective migration both within an intact cell monolayer and in response to cell density gradient such as in an experimental scratch-wound model, where cell-free space is created e.g. by removing cells by making a scratch in the monolayer. These studies have considerably advanced our understanding how such collective migration is organized, e.g. in terms of leadership.

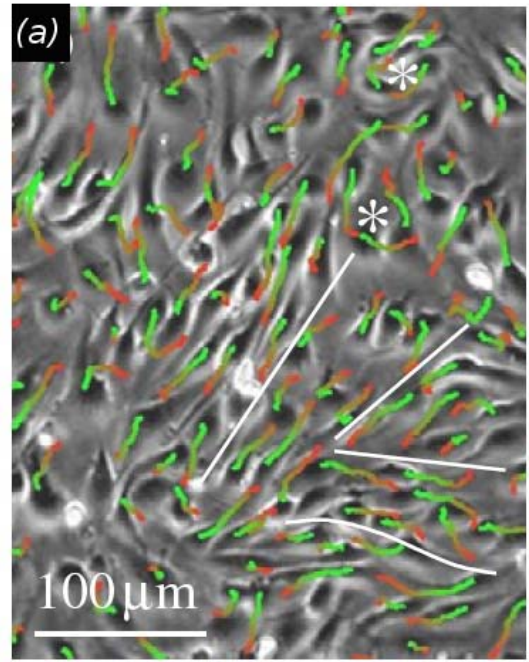

**Fig. 2** Streaming motion of endothelial cells in vitro.
Cell movement within a bovine aortic endothelial (BAEC) monolayer is visualized by cell trajectories in a phase-contrast image with superimposed cell trajectories depicting movements during 1 h. Red-to-green colors indicate progressively later trajectory segments. Adjacent BAEC streams moving in opposite directions are separated by white lines and vortices are indicated by asterisks. From Szabó et al. (2010) with permission of Phys. Biol.

#### Streaming in cell monolayers

In dense monolayers, endothelial cells and various epithelial cells exhibit an intriguing motion pattern, termed 'streaming'. Streaming is a globally undirected but locally correlated motion with emergent internal flow patterns appearing and disappearing at random positions without directed expansion of the whole monolayer. Streaming was observed in the endothelial cell layer lining major blood vessel walls in developing bird embryos[14] and also among immune cells in dense lymph nodes.[15] This form of collective motion, which is different from external chemotactic gradient-driven motility or uncorrelated diffusive motion, was analyzed in cultures and modeled by Czirok and coworkers[4,16] (**Fig. 2**).

#### The role of leadership

The widely accepted approach concerning the nature of migration of groups of cells assumes that "leader cells" situated at the front edge of the group guide the motion of all cells in the group and also provide the necessary traction forces for this. Integration of various intrinsic and extrinsic signals result in the selection of leader cells that polarize and interact with the tissue matrix (see detailed review:[17]).

In experiments with mosaic cultures of wild type vs. specific gene-silenced human endothelial (HUVEC) cells Vitorino et al.[3] have found that the sheet migration evoked by scratch-wound and eventually closing the wound by directed immigration of marginal cells in the cell-free space followed by directed migration of cells localized farther from the boundary is a process regulated in a hypothesized modular way. A functional polarization of cells into leader/pioneer or follower cells occurs at the boundary. Leader cells orient their lamellipodia toward the free space and their motion becomes directed, a process which depends on fibroblast growth factor (FGF) signaling through FGF receptor (FGFR) in the FGFR-RAS-PI3K pathway, but it does not require a concentration gradient of FGF. Migration of the followers several rows behind becomes directed through cell-cell coordination, which depends on the presence of cell surface adhesion molecule VE-cadherin but does not require FGF signaling. Mechanosensing is hypothesized to orient the followers toward the leaders.

The traction forces driving collective migration are generally thought to be exerted by leader cells. However it has been shown[18] that in groups of cultured kidney epithelial (MDCK) cells the traction forces are not exclusively generated by leader cells at the edge but also by cells several rows behind, using cryptic lamellipodia.[19]

Motivated by wound-healing experiments Poujade et al. studied the collective motion of MDCK cell layers triggered by experimental opening up of cell-free surface using a microfabrication-based technique (stencil) without cell damage.[20] This setting with undamaged cells suggests no release of chemical signaling factors at the wound site. In the process of invading the new surface, involving the coordination of many cells distant from the border, they also identified leader cells with directionally persistent motion, active protrusions and focal adhesions at the border. These leaders form fingering instabilities that destabilize the border. Leaders and followers are hypothesized to be coupled by mechanical signaling through the observed cadherin cell-cell contacts among leaders and followers as well as by the multicellular actin cytoskeletal belt at the sides of these fingers. Cell-cell adhesion keeps the monolayer cohesive, which produces long-range correlation in the cell velocity field (**Fig. 3**). Leader cells also originate within the monolayer and brought to the border by streaming flow.

**The role of geometrical confinement**

The impact of geometrical confinement on 2-dimensional collective cell migration has been brought to focus recently by experiments with micropatterned surfaces permitting cell adhesion. In confluent population epithelial cells, collective motion is induced by confinement to areas of physical size below the correlation length of motion measured in the unconfined population. Cell density has a permissive role in this as collective motion does not emerge below confluence.[21] The instructive role of external confinement has been further elucidated by cell velocity field and force distribution mapping experiments. Different in vitro migration modes are induced by 2-dimensional confinement depending on the length scales.

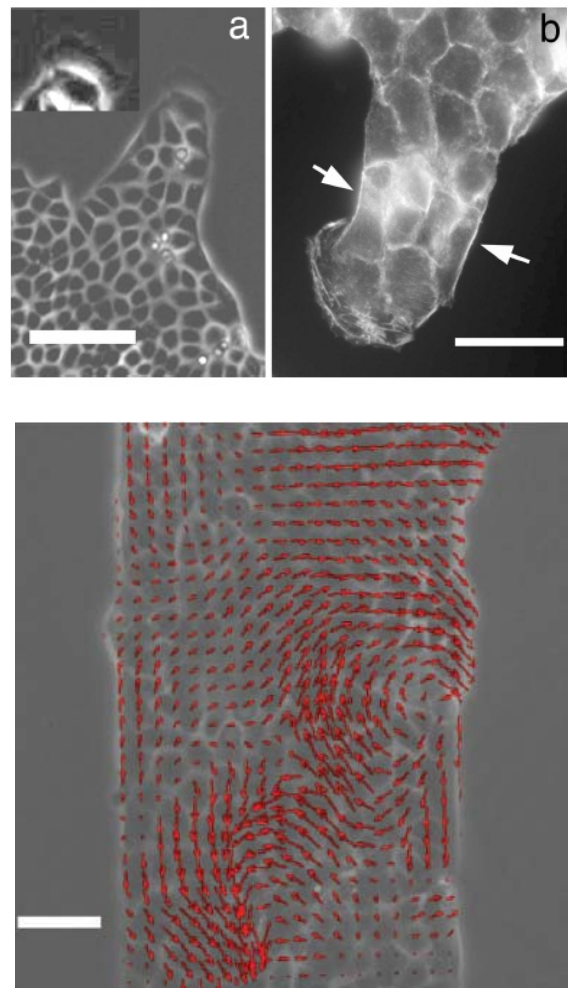

**Fig. 3** Formation of multicellular fingers in cell monolayers.

Upper panel: Micrographs of leader cells 18 h after stencil removal. In each image, a single leader drags a finger. (a) Phase contrast image of a finger preceded by a large leader cell. At the leading edge of this leader there is a very active ruffling lamellipodium (inset: contrast was enhanced on this cell), scale bar: 100 μm. (b) Fluorescence image of the actin cytoskeleton. Particularly visible is the subcortical actin belt along the edges of the finger (arrows), scale bar: 50 μm.

Lower panel: Snapshot of the velocity field 4 h after removal of the stencil. This image was obtained by particle imaging velocimetry. The two vortices are an illustration of how coordinated the flows can be but are not a general feature. Scale bar: 50 μm. From Poujade et al. (2007) with permission of Proc Natl Acad Sci USA.

Epithelial cells confined on narrow strips of width comparable to cell size exhibit a contraction-elongation type of motion with increased migration speed. As a contrast, the same cells on a magnitude wider strips move as sheet under tensile state while exhibiting larger coordination and forming vortices of size comparable to tens of cell size.[22] The role of force transmission through intercellular adhesion contacts has a crucial role in collective migration as coherence is fully abolished by even transient disruption of cell-cell adhesions resulting in cells exhibiting random walk.[23]

# Collective cell motion in vivo

## Collective cell motion in avian embryonic vascular network formation

Early stages of avian embryonic development, drawing the attention of many experimentalists due to its accessibility for observations, is an intermediate state between two and three dimensions. It can be viewed as quasi-two-dimensional because three-dimensional motions take place in an environment confined to essentially two dimensions due to the flattened morphology of the embryo.

One of the spectacular processes of early avian development is vasculogenesis: endothelial cell precursors continuously differentiated in a spatially scattered way in the lateral mesoderm or aggregated in the extraembryonic mesoderm self-assemble into tubes, eventually forming the primary vascular plexus, a polygonal tubular network.[24,25,26,14] Initially scattered precursors divide and locally assemble into vessels or migrate to developing vessels and subsequently move towards the embryonic midline and participate in the formation of large vessels and the heart.

Using transgenic quail embryos (Tg(*tie1*:H2B-eYFP)$^+$) in which all endothelial precursors specifically express a fluorescent marker (YFP) Sato et al.[14] have provided detailed imaging and analysis of endothelial cells' motion in vivo. On the one hand, these cells move passively with gastrulating tissues towards the midline and, on the other hand, they actively move relative to their environment. By the advanced imaging technique, passive motion can be subtracted from overall motion yielding the active motion of endothelial cells. Their active motion does not seem to follow prepatterns in the environment and it is characterized by switching directionality and an apparent attraction to elongated cells and cell chains (also see Reference videos 2 and 3). Endothelial cells eventually assemble into chains of 3-10 cells, giving rise to polygonal tubes (**Fig. 4**).

## Gastrulation of the zebrafish embryo

The universal phenomenon of gastrulation, the formation of the main germ layers of embryos, in various higher animal taxa ranging from fish through amphibians to birds and mammals is an important field where 3-dimensional collective cell migration occurs.

One of the most extensively studied gastrulations is that of the zebrafish, where a crucial phase of the process is the ingression of mesendoderm progenitors from the surface at the mid phase of epiboly, their ingression followed by coherent migration parallel to the surface toward the forming embryonic body axis (**Fig. 5**).

Performing cell transplantation experiments with various genetically modified embryos and cells Arboleda-Estudillo et al.[5] studied the directionality and movement coordination of mesendoderm progenitors. They have found that directional migration of these cells is not a new collective property but already the property of individual cells moving alone.

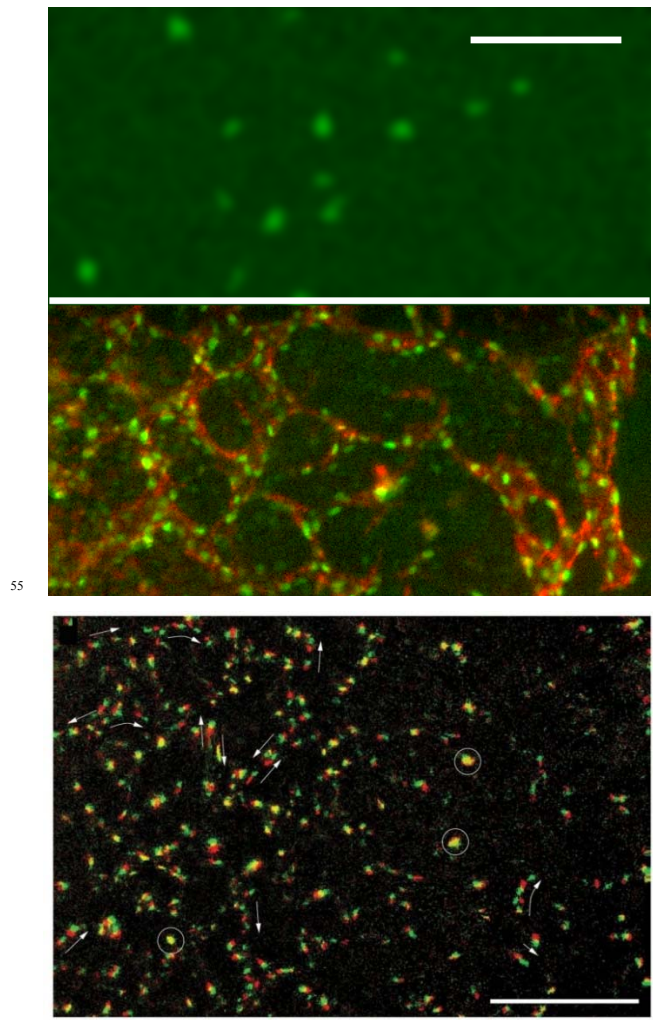

**Fig. 4** Formation of the primary vascular network in the quail embryo.

Upper panel, top: Endothelial cell precursors specifically expressing YFP (green) in their nuclei are scattered in the lateral mesoderm at Hamburger-Hamilton stage 8. Upper panel, bottom: The same part of the embryo 4 hours later. Endothelial cells expressing YFP (green) and also labeled with CyC3-conjugated QH1 antibody (red) against a specific endothelial cell surface marker have self-organized into a polygonal tubular network and the presumptive dorsal aorta (vertical tube at right). The scale bar is 100 μm. Exerted from supplementary videos of Sato et al. (2010) with permission of PLoS One, also see Reference videos 2 and 3.

Lower panel: Cell-autonomous active movement of TIE1+ nuclei, obtained after digitally correcting for the deformations associated with tissue motion in the nascent network during vasculogenesis of the quail. Two consecutive frames, separated by 8 minutes, are shown — the first as red, the second as green. Motile activity is inhomogeneous within the population: some nuclei do not move (appear as yellow, some are marked with circles), while most cells move in a chain-migration fashion (indicated by arrows). At this stage of vasculogenesis, movement directions are highly variable: even in the same vascular segment, groups/chains are seen moving in opposite directions. Scale bar: 200 μm. From Sato et al. (2010) with permission of PLoS One.

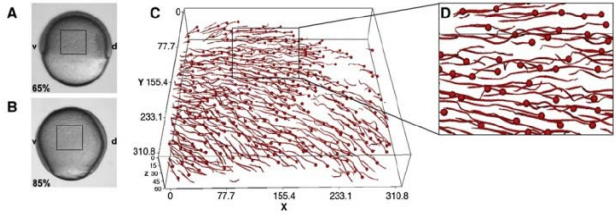

**Fig. 5** Movement of lateral mesendoderm cells in wild-type embryos. (A and B) Brigth-field images of an embryo at the beginning of gastrulation (6.5 hours postfertilization [hpf]; A) and at midgastrulation (8.5 hpf; B) Boxes outline the imaged region in (C). (C) Trajectories of mesendoderm progenitors during midgastrulation stages. Nuclei were tracked and the endpoint of each track is indicated with a sphere. The box depicts the magnified region shown in (D). Embryos were imaged by two-photon excitation microscopy from 6.5 to 8.5 hpf. Animal pole is to the top and dorsal is to the right. From Arboleda-Estudillo et al. (2010) with permission of Curr Bio.

Nevertheless the collective migration of mesendoderm cells is impaired and becomes less directed if cell-cell adhesion is defective, as it was shown by modulating cell-cell adhesion strength through the modulation of E-cadherin expression, the key adhesion molecule in mesendoderm cells (also see Reference videos 4 and 5). To analyze the contribution of cell-cell adhesion to collective mesendoderm migration they used a numerical simulation.

Other aspects of the collective migration of mesendoderm cells in gastrulating zebrafish embryos were studied recently.[27] Single mesendoderm cells or small groups were transplanted ahead of the advancing prechordal plate (the front part of the ingressing mesendoderm), an area most likely permissive for their directional migration. These single motile cells or small groups, however, failed to migrate in the right direction toward the animal pole but stayed in position or migrated backward until joining the advancing prechordal plate where they were quickly re-oriented taking the direction of the prechordal plate through active motion, i.e. they were not dragged or pushed passively. Cell-cell interactions and contact with the endogenous prechordal plate are required to orient the motion of these cells in which the major components are E-cadherin-based adhesion, cell polarity defined by the Wnt-Planar Cell Polarity signaling pathway and directed cell protrusion activity regulated by Rac1 GTP-ase.

Mechanosensing the tension gradient developing within the advancing prechordal plate by an intrinsic mechanism without extrinsic cues is hypothesized to account for this self-organization, a mechanism yet to be explored experimentally.

Various aspects of force generation and regulation in morphogenesis are discussed in an excellent recent review.[28]

### Collective migration of the posterior lateral line primordium of the zebrafish

The development of the lateral line organ in the zebrafish is a series of 3-dimensional collective migration events that are both well characterized biologically and integrated in a computational model (**Fig. 6**).

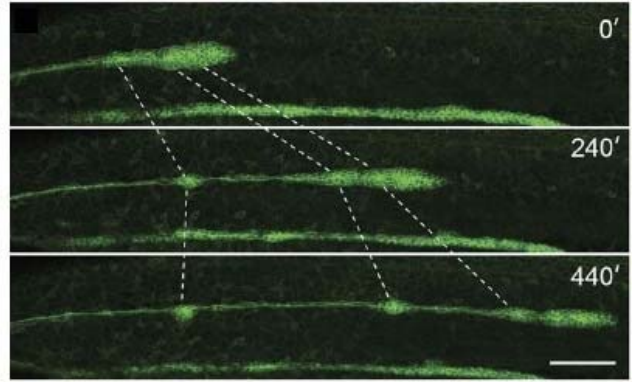

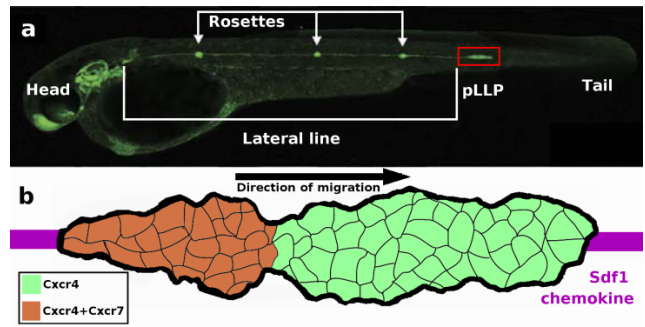

**Fig. 6** The posterior lateral line primordium couples collective migration to differentiation.

Upper panel: An overview of a time-lapse movie showing 10 hr of lateral line morphogenesis with Claudin B-GFP. The lateral line primordium migrates at a speed of ~66 μm/h at 25°C. Forming neuromasts at the trailing edge (dotted lines) decelerate, causing the tissue to stretch, before being deposited. The scale bar is 100 μm. Also see Reference video 6. From Haas and Gilmour (2006)[6] with permission of Dev Cell.

Lower panel, (**a**) Microscopic image of the zebrafish embryo at 42 hpf. The posterior lateral line primordium (pLLP, red box) and rosettes are visible due to Claudin B-GFP marker. Modified from Haas and Gilmour (2006)[6] with permission of Dev Cell.

(**b**) Schematic image of the pLLP corresponding to the area highlighted in red box in (a). The primordium migrates along the Sdf1 chemokine prepattern (purple stripe), detected by CxCr4 receptor (green). The trailing region of the primordium also express Cxcr7 receptor (overlap of the two receptors is seen in orange).

During organogenesis, the primordium of the lateral line organ, a series of mechanosensory hair cell organs, differentiates from neurogenic placodes on both sides of the embryo's head region. The posterior lateral line primordium (pLLP), which is a cohesive mass of more than 100 cells, then migrates as one cohort along a defined path at the side of the embryo while depositing clusters of neuromast cells transforming into sensory epithelial cells forming a series of connected groups, termed rosettes, constituting the lateral line organ. The migration of the primordium is completed in less than 12 hours (see Reference video 6).

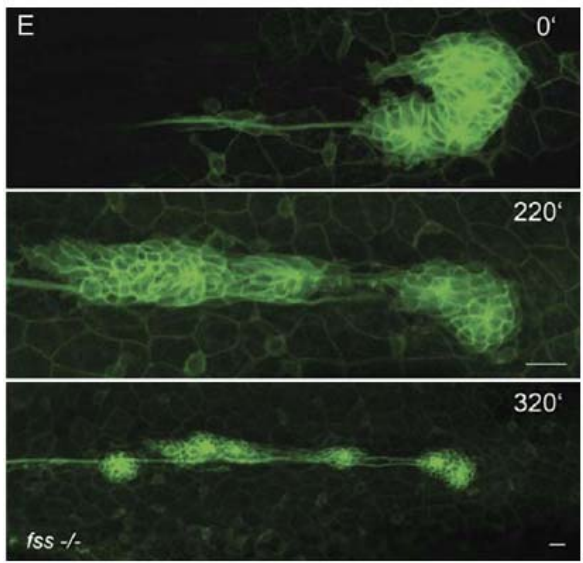

**Fig. 7** Overview of a time-lapse movie showing the lateral line primordium undergoing a "U-turn" maneuver.

The upper ''start'' panel shows a rounded primordium; a small group of cells projects backward, causing the tissue to rotate. Once this ''U-turn'' is complete, the primordium readopts its normal polarized morphology and migrates at normal speed in the reverse direction and even deposits a proneuromast. Also see Reference video 7. From Haas and Gilmour (2006) with permission of Dev Cell.

The path followed by the primordium is defined by a chemokine, stromal-derived factor 1 (Sdf1a, also termed Cxcl12a), expressed by the surrounding myogenic tissue in a stripe pattern, detected by the primordium through expression of the receptor CxCr4b. Although most cells of the primordium express Cxcr4b, only few cells at the leading tip activate the receptor to direct the polarity of the whole group, hence acting as leader cells. Genetic mosaic experiments have revealed that cells with mutant receptor are specifically excluded from the leading edge implying that adequately functioning CxCr4b receptor is required for becoming a leader cell, whereas it is not required for being a follower cell. Here, mechanical force exerted by leaders on followers through the N-cadherin cell-cell contacts is hypothesized to guide followers.

In the absence of either of the receptors, Cxcr7 or Cxcr4b, or their ligand, Sdf1a, the migration of the primordium is seriously defective. Cxcr7 is thought to be required at the rear to ensure persistent forward migration of the whole primordium while regulating the halting and deposition of rosettes through an intracellular signaling differing from that of Cxcr4b[6,29] (a detailed review is also available:[11]).

As the primordium advances, a fibroblast growth factor, FGF10, expressed in discrete spots by the adjacent tissue induces follower cells to adopt an epithelial cell fate and generate the rosette-like structure. Simultaneously, the trailing region of the primordium slows down and halts causing elongation of the primordium followed by seceding of the rosette. This process correlates with the presence of another receptor of Sdf1a, Cxcr7, expressed only by follower cells mainly at the trailing region of the primordium while there is a large overlap with Cxcr4b expression.

Experimental truncation of the Sdf1a stripe can cause a 180 degree turn of the entire migrating primordium followed by migration in reverse direction and normal depositing of neuromasts (see Reference video 7). This suggests that there is no polarized distribution or long-range concentration gradient of the chemokine guidance cue, but polarization rather lies in the organization of the migrating primordium itself[6] (**Fig. 7**).

By establishing a novel readout of chemokine ligand activity based on visualizing and measuring the turnover of the ligand binding receptor, using a tandem fluorescent protein timer (lifetime tFT) method, Gilmour and coworkers have recently provided direct evidence for the self-generation of chemokine gradient by the migrating collective itself.[30]

The Sdf1a ligand concentration-decreasing activity of Cxcr7 receptor, expressed at the rear of the primordium, is sufficient to generate a gradient of chemokine activity across the primordium's whole length, dispensing the necessity for pre-existing long-range gradients that may have spatial limitations.

**Collective chemotaxis: migration of neural crest cells in embryonic development**

During embryonic development of vertebrates, two parallel stripe-shaped areas at the borders of the neural plate on both sides of the forming neural tube detach from the neuroectoderm through an epithelial-to-mesenchyme transition process called delamination and eventually form the neural crest (NC). It is a neurogenic tissue, which becomes segmented, giving rise to various elements of the peripheral nervous system. Additionally, many neural crest cells migrate long distances from their original site at the dorsal midline towards the ventral regions and participate e.g. in the formation of the adrenal gland while others colonize to the forming dermal tissue as pigment cells. This ventral-directed migration of dynamically reshaping cell clusters, streams or cell chains is known to be instructed by several diffusible chemotactic agents (attractants and repellents) produced externally while coherent directional migration is controlled by interactions among cells. Specifically, N-cadherin-mediated contact inhibition of locomotion (CIL), a short range repulsive interation among neighboring cells facilitates the growth of protrusions at non-inhibited free surfaces leading to directional polarization and higher directional persistence of migration.[31] Cohesion of the group is maintained by longer-range mutual attraction (coattraction) of cells through mutual production and binding of the ligand complement fragment C3a by its receptor C3aR. The directional polarization induced by CIL is stabilized and amplified by the chemokine ligand Sdf1, bound by its receptor Cxcr4, while the migrating collective can functionally differentiate into leaders and followers with dynamic shuffling of roles and the groups themselves can split and reassemble.[32] Compared to single NC cells, a group of various number of NC cells can more efficiently migrate towards the chemokine by such 'collective chemotaxis'.[33,34]

**Collective migration in branching morphogenesis: development of the trachea network**

Branching morphogenesis is a form of collective cell migration playing pivotal role in the formation of various structures in embryonic morphogenesis or tissue development or regeneration in adults.

The tracheal system of the fruit fly, D. melanogaster, and the vascular system of birds and mammals are two exemplary areas where branching morphogenesis leading to the formation of a tubular system is studied. A common theme to all these tubular systems is their branched and hierarchical nature. The morphological similarity among various tubular systems is related to similarities between the signaling pathways and biophysical characteristics controlling their branching and growth (for detailed review see:[35]).

Experimental work with embryonic model systems led to the identification of ligand-receptor pairs involved in the persistent directional migration and guidance of cell groups forming these structures. They have also improved our understanding how the leader-follower organization of groups is determined by initial symmetry breaking events mediated by other ligand-receptor pairs.

The development of the tracheal system in the fruit fly Drosophila melanogaster takes place without cell proliferation and eventually the collective migration of 10 groups each consisting of ~80 ectodermal cells is responsible for its formation. Tip cells differentiate as leader cells of the group and produce dynamic cytoskeletal protrusions, then form the primary branches by migrating toward a fibroblast growth factor (FGF) source produced in defined patches by cells surrounding the group. The tip cell prevents its neighbors from becoming leaders in a process called lateral inhibition. The molecular mechanism of adopting a tip cell fate was studied by Ghabrial and Krasnow[7] and reviewed by Schottenfeld et al.[36] Initial slight differences in FGF receptor signaling are amplified by positive and negative feedback loops and eventually lead to increase in the expression of Notch receptor ligand Delta in the leader tip cell. Delta activates Notch in the neighboring cell which eventually downregulates the FGF receptor pathway and Delta expression in the neighboring cell thus making it less responsive to the FGF signal and becoming a follower stalk cell.

The dynamics of cell fate segregation through lateral inhibition by the Delta/Notch system was studied using mathematical models.[37,38] Analysis of a model of a lateral inhibitory system along with a spatial gradient of its input stimulus has revealed that such a system mainly contributes to the robustness of tip-cell selection when the input signal includes random noises, which is frequently the case in complex developmental processes. It has also been shown that lateral inhibitory regulation works more robustly in tip-cell selection than self-inhibition, an alternative means of inhibitory regulation.

**Collective migration in branching morphogenesis II: development of the vascular network**

A very intensively studied field of branching morphogenesis is vascular sprouting and the formation of vascular networks. Avian embryos have become the model organisms for vascular research due to their ease of accessibility and because of similarities with the vascularization of murine embryos, suggesting a generic mechanism shared by warm-blooded animals.[24,39]

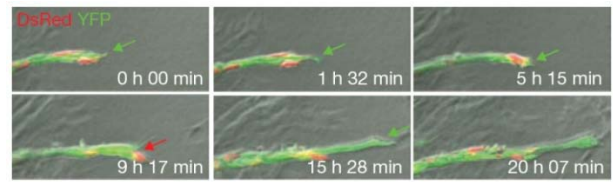

**Fig. 8** Dynamic observations of tip cell shuffling in sprouting angiogenesis.
Time-lapse microscopy images of chimaeric embryoid bodies of wild-type cells expressing DsRed (red) or YFP (green). Red arrow indicates when a green cell is overtaken by a red. From Jakobsson et al. (2010).with permission of Nat Cell Biol.

During embryonic development of warm-blooded animals the first phase of vascular network formation is termed primary vasculogenesis in which endothelial precursors randomly differentiated in the lateral mesoderm self-assemble by active motion into a polygonal network, yet void of fluid. The second phase is termed angiogenesis when this initial vascular network already carries blood and it is further reshaped by vessel sprouting, fusion or withdrawal on demand by surrounding tissues and hemodynamic forces. Angiogenesis, essentially the outgrowth of new vessels from existing vessels, then occurs throughout life as endothelial cells are capable of developing networks in several modes in various biological conditions and tissue environments.

Candidate mechanisms for vascular patterning include: guidance by pre-pattern, contact guidance by extracellular matrix (ECM) and mechanosensing, guidance by interactions modifying the ECM (referred to as 'ECM memory') and guidance by chemotactic gradients. A very detailed review on vascular patterning mechanisms has been published recently.[16]

In vascular sprouts, the endothelial cells are guided by a single tip cell protruding actin-rich filopodia, followed by a multicellular stalk of endothelial cells, connected by vascular endothelial cadherin (VE-cadherin) at cell-cell junctions successively forming the inner lumen of the new vessel.

As the initial step of vascular sprouting a differentiation step to become leader tip cell vs. follower stalk cell occurs similarly as in tracheal morphogenesis. In endothelial cells the vascular endothelial growth factor (VEGF) and subsequent Delta-Notch signaling axis determines leader and follower cell fate by lateral inhibition.

Jakobsson et al. studied the molecular mechanism of tip cell selection in angiogenesis in the retina and in embryoid bodies.[8] They have found that endothelial cells dynamically compete for the tip cell position through relative levels of VEGF receptor (VEGFR) subtypes 1 and 2. Dynamic position shuffling of tip cells and stalk cells has been observed in experimental sprouting assays (**Fig. 8**).

Differential VEGFR levels modulate the expression of the Notch ligand Delta (Dll4) activating Notch in the neighboring cell, which in turn influences the expression level of VEGFR subtypes. Cells with lower VEGFR1 and higher VEGFR2 levels are more likely to take and maintain the leading position.

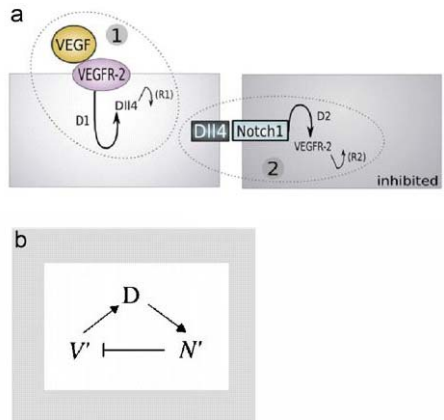

**Fig. 9** (a) The two pathways involved in notch-mediated tip cell fate determination. D1 and D2 are transcriptional delays. R1 and R2 are recovery delays representing the time it takes before gene expression returns to normal. d and s represent expression levels in response to receptor activation or loosely, transcription factors. (b) The pathway as a negative feedback loop, active VEGFR-2 (V0) induces Dll4 (D), which increases active Notch1 (N0) leading to VEGFR-2 inhibition. From Bentley et al. (2008) with permission of J Theor Biol.

Based on data from in vitro and in vivo sprouting experiments with genetic chimaeras Bentley et al.[40] developed a hierarchical agent-based computational model for the simulation of sprouting in uniform and gradient distribution of VEGF. Simulation results show that Notch-dependent regulation of VEGFR2 can function to limit tip cell formation from the stalk in a competitive way (**Fig. 9**).

**Vasculogenesis by biophysical mechanism**

Vascular sprouting can be viewed as an emergent process governed, at least in part, by biophysical rules influencing the motion of cells involved.

An in vitro model system where primary vasculogenesis can be studied experimentally is the allantois formed by the lateral extraembryonic mesoderm in both birds and mammals.[41] Within the allantois, vasculogenic cell aggregates, termed blood islands, give rise to sprouts eventually forming a vascular network. Endothelial cells are also capable of forming networks in various in vitro systems, such as 3D collagen hydrogels, where environmental or genetic pre-patterns are obviously missing.[42,43] After dynamic competition for tip cell position, angiogenetic sprouts are led by very motile tip cells while similarly motile stalk cells are recruited from aggregates and follow the tip cell while occasionally overtaking it.

It is tempting to think that stalk cells are passively dragged by the tip cell but if so the elongation of the sprout would be limited because the cadherin-mediated cell-cell adhesions, shown to be analogous to surface tension of liquid droplets, would not be able to stabilize the structure beyond a critical length. Due to the Plateau-Rayleigh instability a surface-tension stabilized structure, such as a liquid jet, will break up into drops when its length exceeds its circumference. Sprouts grow beyond this length indicating that stalk cells actively move within an expanding sprout following some sort of a guidance mechanism. To search for a potential guidance mechanism to recruit stalk cells in the expanding sprout Szabo et al. studied sprouting in a simplified in vitro system without chemokines.[44] They have demonstrated that various non-endothelial cell types can also exhibit the sprouting behavior on 2-dimensional surfaces, suggesting a generic mechanism.

**Vasculogenesis by chemotaxis**

Vascular sprouting can also be viewed as a process guided by autocrine chemotactic signaling where the process relies on the secretion of a diffusible chemotattractant morphogen by cells.[45,46,47]

In avian embryonic vasculogenesis, however, the chemoattractant VEGF165, which likely fits in the model, is produced throughout the embryo and overweighs the low autocrine production, if any, by endothelial cells. The same applies to in vitro 3D collagen invasion assays where endothelial cells readily form sprouts and network in the presence of high concentration of exogenous VEGF in the medium.

These contradictions can be overcome if it is assumed that VEGF binds to the extracellular matrix (ECM) while endothelial cells secrete a proteolytic agent releasing the ECM-bound VEGF creating a local gradient of the "bioavailable" VEGF in the microenvironment of endothelial cell aggregates, pointing towards the aggregates. Such a mechanism has not yet been validated experimentally mainly owing to difficulties in visualizing or measuring morphogen gradients.

A recently emerging hypothesis based on the effect of a diffusible inhibitor also attempts to solve the above contradictions.[16] Experiments with diffusible VEGF receptor (VEGFR1) secreted by endothelial cells show that lack of this secreted receptor severely compromises vascular sprouting, whereas exogenous soluble VEGFR1 production by endothelial cells in the vicinity of emerging sprouts can rescue sprout formation and elongation.[48,49] Based on these findings it can be hypothesized that diffusible VEGFR1 secreted by endothelial cells binds and sequesters the otherwise abundant VEGF in the vicinity endothelial cells, creating a VEGF gradient pointing away from these cells.

**Pattern formation by collective segregation of cells**

An interesting field where collective cell motion is involved is the spatial pattern formation by different cell types through the process termed segregation (or sorting). Patterns can form as a response of cells to external guidance cues such as morphogens or chemotactic substances or as a process where instead of external cues the local cell-cell interactions and inherently different mechanical or motility characteristics of cell types give rise to various multicellular patterns by physical segregation of the cell types.

These segregation events bear much significance in embryonic development of higher animals where differentiation, pattern formation and cell motion take place simultaneously. A recent review summarizes several cell segregation phenomena and corresponding computational models.[50]

**Pattern formation in cell monolayers in vitro**

Basic drives and mechanisms of pattern formation events taking place e.g. during embryonic development can be studied in simplified experimental systems where complexity is reduced and the events are more accessible for quantitative analysis.

In 2-dimensional co-cultures of adherent cells on a rigid substrate Mehes et al. studied the dynamics of segregation of two initially mixed cell populations into distinct clusters by cell migration in an environment lacking pre-defined external cues.[51] They have found that segregation into large multicellular clusters is facilitated by collective effects in cell motion such as an increase in the directional persistence of constituent cells. The growth of such multicellular clusters by consecutive fusion of smaller clusters follows algebraic scaling law with characteristic exponents depending on the collective effects (**Fig. 10,** also see Reference video 8).

The growth exponent values measured in this cell culture system with self-propelled collective motion exceed the exponent values resulting from computer simulations with diffusively moving segregating units detailed in a report by Nakajima and Ishihara.[52]

**Pattern formation by segregation in vivo:**

**gastrulation and tissue organization**

Three-dimensional segregation of cell populations is most prominent during gastrulation, the early phase of embryonic development resulting in the formation of main germ layers that later on give rise to all tissues. Gastrulation is a spectacular event under the microscope involving collective motion of large number of cells, but although gastrulation events have been known since early embryonic works at the beginning of the 20th century, the basic mechanisms that provide for both its accuracy and robustness are just being uncovered. Segregation of cell populations with different cell fates into distinct domains is governed by their mechanical properties and active motion, and it is an important driving mechanism of gastrulation and tissue organization. Segregation is also important in other embryonic processes ranging from blastocyst formation to somitogenesis in vertebrates.

Cell segregation was first demonstrated by the experiments of Townes and Holtfreter in which presumptive neural and epidermal cells were isolated from amphibian gastrulas, subsequently they were mixed and they autonomously sorted into separate tissues.[53] In similar early experiments, mixed cells isolated from the adult Hydra were shown to segregate and form separate tissues.[54]

Segregation of various cell types in 3-dimensions was studied in several works[55,56,57,58,59] aiming to explain the observed configurations of segregated domains, typically the envelopment of one cell type by the other, evolving from an initial mixture of cells. These in vitro segregating systems are considered to be analogous to non-mixing liquids and their segregation is shown to be driven by differences in tissue surface tension (TST) of the constituent cell types.[55] Several works tested the contribution of cell-cell adhesion[57,59] and cell cortex tension[58,60] to TST.

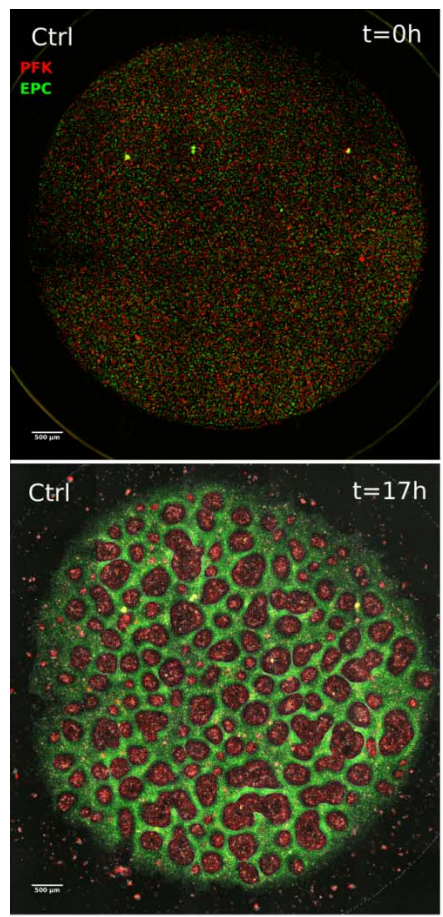

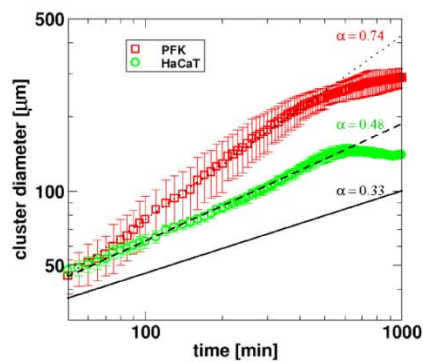

**Fig. 10** Dynamics of 2-dimensional segregation of keratocytes in culture.

Upper panel: Segregation in mixed co-cultures of primary goldfish keratocytes (PFK, red) and EPC fish keratocytes (EPC, green), consisting of >250 000 cells. Top panel shows initial stage after cell attachment, middle panel shows final stage after 17 hours of cell migration. Scale bar is 500 μm. Also see Reference video 8.

Bottom panel: Average cluster diameter growth curves calculated from experiments with primary goldfish keratocytes (PFK) or human keratocytes (HaCaT). Exponent values obtained from fitting straight line segments to the experimental curves are shown. Cluster growth curve of simulated segregation of cells without collective motion characterized by exponent value α = 0.33 is shown for reference (black solid line). From Mehes et al. (2012) with permission of PLoS One.

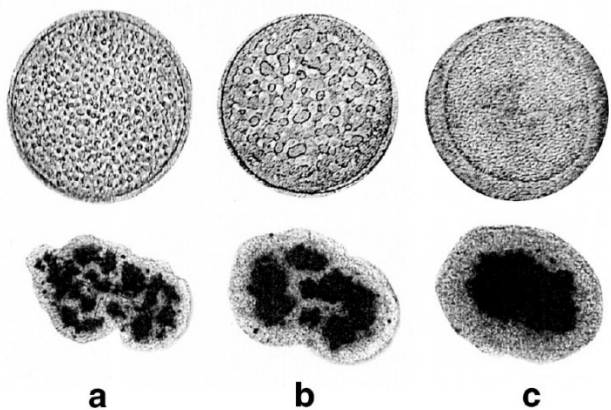

**Fig. 11** Gas and liquid phase ordering and segregation of retinal cells.

Upper panel: Gas and liquid phase ordering in SF6 under reduced gravity, after a thermal quench of 0.7 mK below the critical point (45.564 C). Gas and liquid eventually order with the liquid phase wetting the container wall and surrounding the gas phase, corresponding to wall-liquid interfacial tension < wall-gas interfacial tension. *a*, *b* and *c* correspond to 120 s, 275 s and 3960 s after quench, respectively.

Lower panel: Sorting out of chicken embryonic pigmented epithelial cells (dark) from chicken embryonic neural retinal cells (light). The average aggregate size is 200 μm. At the end of sorting, neural retinal cells preferentially wet the external tissue culture medium surrounding the aggregates. Medium-neural retina and medium-pigmented epithelium interfacial tensions are 1.6 dyne/cm and 12.6 dyne/cm, respectively. *a*, *b* and *c* correspond to 17 h, 42 h and 73 h after initiation of sorting, respectively. From Beysens et al. (2000) with permission of Proc Natl Acad Sci.

**Three-dimensional segregation experiments**

The dynamics of growth of segregated domain size in 3-dimensions was studied by Foty et al.[55] using mixed cultures of embryonic pigmented epithelial and neural retinal cells, which segregated and formed enveloped structures over time in a configuration determined by surface tensions of the cell types. As a comparison, the segregation of gas and liquid phases was studied under microgravity resulting in similar segregated configuration determined by surface tension (**Fig. 11**).

Authors have found that the size of segregated cell domains and segregated gas/liquid domains both increase linearly in time.

In a study quantifying the adhesive and mechanical properties of zebrafish germ line progenitor cell types Heisenberg and coworkers investigated the role of tensile forces in cell segregation.[57] Using single-cell force spectroscopy they have measured the cell-cortex tension of these cell types (ectoderm, mesoderm and endoderm) while specifically interfering with actomyosin-dependent cell-cortex tension.

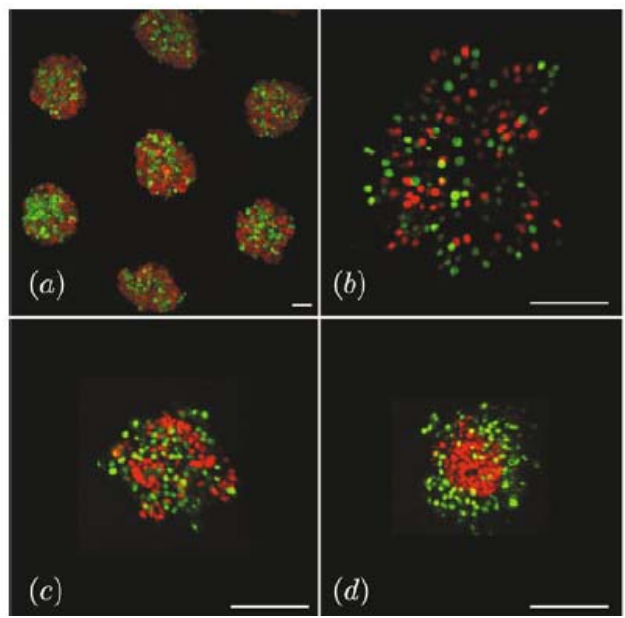

**Fig. 12** Imaging data for different time points in a segregation experiment with zebrafish ectoderm and mesoderm cells in culture.

(a) Micro-molds are used to isolate small populations of ecto- and mesoderm cell mixtures labeled fluorescently with red and green nuclei, respectively.

(b) Initial images show homogeneously mixed cells distributed throughout the mold. (c) Cells aggregate together on a time scale of roughly 100 minutes. (d) Imaging after sorting clearly shows the segregation of the two cell populations. Scale bar = 100 micron. From Klopper et al. (2010) with permission of Eur Phys J E Soft Matter.

Performing segregation experiments using cell types with altered myosin activity they have demonstrated that differential actomyosin-dependent cell-cortex tension is required and sufficient to direct the segregation of cell types and determines the final configuration of the segregated domains.

The dynamics of 3-dimensional segregation of mixed germ line progenitors of the zebrafish was studied by Klopper et al.[61] As segregation proceeds in this system, the domain consisting of mesoderm cells gradually engulfs the ectoderm domain, which eventually takes the inner position (**Fig. 12**). Authors have monitored the dependence of the local segregation order parameter on system size and found algebraic scaling and different characteristic exponent values for enveloping and engulfed cells.

In a similar in vitro system composed of two mixed epithelial cell types suspended in micro-molds, Vicsek and coworkers have recently studied the dynamics of 3-dimensional segregation[49] (see Reference video 9). In their experiments the forming domains are adjacent and unlike zebrafish germline progenitors there is no engulfment of one domain by the other.

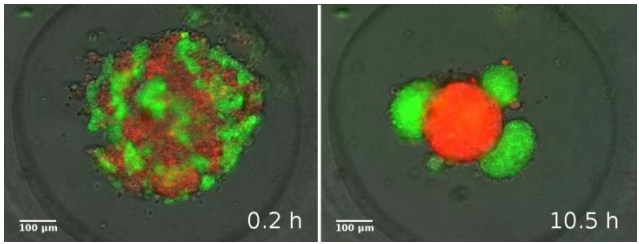

**Fig. 13** Snapshots from a segregation experiment with two keratocyte types in culture. Left: Initial mixture of primary goldfish keratocytes (stained red) and EPC fish keratocytes (stained green) in a micro-mold after onset of segregation. Right: Homotypic cell clusters formed through segregation. Also see Reference video 9. From Mehes and Vicsek (2013) with permission of Complex Adaptive Syst Model.

It was also found that the growth of segregated domain size follows algebraic scaling law and it is fast, typically completed within 6 hours (**Fig. 13**). These observations are in harmony with simulations of Mones et al.[62] but in contrast with earlier simulations of Chaté and coworkers[63] that suggest a much slower process (see **Fig. 25**).

Pattern formation by segregation is a process that is not confined to embryonic development. In a recent publication Inaba et al.[64] studied the formation of skin pigment patterns in the adult zebrafish. They have demonstrated that segregation of the two pigment cell types eventually forming the stripe pattern is governed by their short-range repulsive electric interactions that spatially orient their migration.

**Emerging hypotheses**

Two opposing hypotheses have been developed for explaining the origin of tissue surface tension, TST, the main drive of collective cell segregation. One is the differential adhesion hypothesis (DAH), developed by M. Steinberg[65,66,67,68] postulating that tissue surface tension is proportional to the intensity of adhesive energy between point object cells. This hypothesis was elaborated in extensive modeling approaches by J. Glazier.[69] Experimental studies showed that TST is proportional to cadherin levels.[57]

The other hypothesis is the differential interfacial tension hypothesis (DITH), developed by Harris[70], Brodland[71,72] and Graner[73], postulating that tissue surface tension arises from cortical tension of individual cells generated by actomyosin contractility, while a cell's mechanical energy changes with cell shape. This model was also supported by experimental data on cell cortex tension and TST.[58]

A model integrating cell-cell adhesion and contractility of cell interfaces in the generation of tissue surface tension, the driving force of cell segregation and tissue spreading, was provided by Manning et al.[74] This model specifies an explicit relationship between surface tension and the ratio of adhesion ($\gamma$) to cortical tension ($\beta$). Surface tension exhibits a crossover at $\gamma/\beta \sim 2$ from adhesion-dominated behavior (DAH) in the regime of $\gamma/\beta < 2$ to a dependence on cortical tension and other mechanical effects in the regime of $\gamma/\beta > 2$.

Experimental proof on the relative weights of adhesion and cell cortex tension in controlling cell-cell contact formation in zebrafish germ layer progenitors and determining the experimentally measurable separation force between cell pairs was provided by Maître et al.[75] Cells are described as fluid objects with viscoelastic cortex under tension and adhesive bonds maintaining cell-cell contacts. Contact expansion is controlled by cell cortex tension at the contact, generated by myosin activity, while adhesion by cadherin molecules (membrane-spanning adhesion molecules) mechanically couple the adhering cells, and such coupling is limited by cadherin anchorage to the sub-membrane cortex. Contact formation is the result of active reduction of cell cortex tension at cell-cell interface, which leads to decrease in cell-cell interface tension, while cell cortex tension at the cell-medium interface will not decrease, accounting for maintained TST. Adhesion is shown to have little direct function in contact expansion. Considering the typical cadherin density, the adhesion energy per unit area of the cell surface ($\sim 1 \times 10^{-7}$ N/m) is several orders of magnitude lower than typical TST measured in cell aggregates (being in the order of $1 \times 10^{-3}$ N/m).[94-76] The main drive of cell contact formation and segregation is actomyosin-dependent cortex tension rather than adhesion energy.

A recent review emphasizes the role of boundary cells in TST as they can actively change their mechanical properties generating different cortical tensions along their internal and external interfaces. Such 'mechanical polarization' is suggested to exert the same net mechanical effect on the tissue as if extra adhesion was introduced among all cells and it is hypothesized to dominate TST instead of the mechanical energy of adhesive bonds.[75] Strong apical-basal actin polarization was shown in surface cells in zebrafish embryonic explants.[77] Considering the low adhesion energy of cadherins, the findings that TST is proportional to the number of surface cadherins[56] can also be interpreted in a way that it is actually signaling through more cadherins leading to increased actomyosin contractility and resultant cell cortex tension which generates higher TST.

## Conceptual interpretations

When attempting to put the relatively new topic of collective cell migration into a wider perspective we shall consider three major aspects of these phenomena. i) Collective motion can be looked at as one of the simplest manifestations of collective behavior. ii) Although a general theoretical framework for such emergent processes as the coherent motion of cells is still lacking, a classification of the collective motion patterns can be a helpful tool in interpreting the various related phenomena. iii) By using a system of equations the description is, on one hand, elevated to a quantitative level and on the other hand since the same equations can be applied to rather different systems, this also indicates the universal emergent features of the collective motion of cells.

**Emergence and collective behavior**

Collective behavior applies to a great many processes in nature, which makes it an extremely useful concept in many contexts. Examples include collectively migrating bacteria, insects or birds, simultaneous stopping of an activity (e.g., landing of a flock of pigeons) or phenomena where groups of organisms or non-living objects synchronize their signals or motion, e.g. think of fireflies flashing in unison or people clapping in phase during rhythmic applause. The main features of collective behavior are that an individual unit's action is dominated by the influence of its neighbors, the unit behaves differently from the way it would behave on its own; and that such systems show interesting ordering phenomena as the units simultaneously change their behavior to a common pattern.

Over the past decades, one of the major successes of statistical physics has been the explanation of how certain patterns can arise through the interaction of a large number of similar units. Interestingly, the units themselves can be very complex entities too, and their internal structure has little influence on the patterns they produce. It is much more the way they interact that determines the large-scale behavior of the system. Extremely complex units (e.g. cells, cars, and people) can produce relatively simpler patterns of collective behavior because their interactions (or behavior from the point of view of the outside world) can have a form that is much simpler than the structure of a unit itself.

### Classes of collective migration of cells

From a general viewpoint, collectively moving entities may exhibit only a few characteristic motion patterns. Some of these are listed with particular examples in the section on the main types of collective cell motion. Modeling and simulational approaches use the notion of self-propelled particles in order to interpret the various collective motion patterns occurring in a wide range of systems containing units that tend to move with an approximately constant velocity and interact through relatively simple forces (repulsion, alignment, etc.). The studies have shown that there are only a few possible states of such systems. The list includes the following relevant cases: i) disordered motion (the direction of motion of the units is not correlated), ordered motion (even distant units move in an approximately same direction), iii) "turbulent motion" (there is local order but it is lost on a scale much larger than the size of the units), iv) "steams" of units flowing opposite to each other and finally, v) "jamming" when the restricted volume and mutual "pushing" of the units results in a highly strained, locally fluctuating but globally not moving groups of particles.

Most of the observations presented above can be looked at as either analogous to one of the above general classes or being a combination of two of them.

### Interpreting collective motion of cells in terms of models/equations

In the next section of this Review we shall discuss two types of models both involving equations for the positions and the velocities of the cells. First we shall consider the simplest or "minimal" models which possess simple rules required for the emergence of collective motion. The second type of models takes into account a few further interactions, already somewhat specific to the particular experimental situation. We shall not discuss the third approach which comprises systems of partial differential equations (continuum approach) because this framework is very theoretical.

However, all three approaches lead to collective motion patterns similar to many of those observed in experiments. We shall show that indeed, equations can be used to interpret phenomena like, for example, the faster segregation of cells as a result of collective effects. Since the above mentioned equations contain only a couple of terms they cannot account for the large number of potential factors that may influence the detailed, actual motion of a cell. This can be done because details "average out" when the behavior of the whole is considered. As a consequence, it is expected that the collective motion of units has characteristic features typical for many different systems. From the point of statistical physics these could be considered as "universality classes" or major types of behavioral patterns. Observing and interpreting these patterns and their relationship to the systems which exhibit them is likely to lead to a unified picture or, in an ideal case, to the discovery of a number of basic relations or "laws" for the collective motion of cells in various biological processes.

## Quantitative models

Interactions of various moving cells with their heterogeneous environment, such as in wound healing, embryonic morphogenesis, immune reactions and tumor invasion have been investigated using mathematical models (for review see: [78]). As an example, a lattice-gas cellular automaton model has been used for modeling in vitro glioma cell invasion and it allows for direct comparison with morphologies and mechanisms of invading collectives.[79] Computational cell biology, an emerging interdisciplinary field, attempts to mediate among several scientific communities investigating various aspects of cell motion (for review see: [80]).

### Simplest models

In this section we first quickly review the basic computational models for the swarming behavior in general and for the collective motion of cells as well. In the subsequent sections the more detailed models that are used for explaining specific cellular phenomena will be introduced as well.

In order to establish a quantitative interpretation of the behavior of large flocks, or cell populations in this particular field, in the presence of perturbations, a statistical physics type of approach was introduced by Vicsek and co-workers.[81] In this cellular-automaton-like approach of self-propelled particles (SPPs), the point-like units move with a fixed absolute velocity, $v_0$, and assume the average direction of others within a given distance $R$, characterized by its angle $\theta_i$:

$$\text{x}\, \theta_i^{t+1} = \arg\left[\sum_{j \sim i} \vec{v}_j^{\,t}\right] + \eta \xi_i^t, \qquad \text{(Eq. 2)}$$

where $\vec{v}_i^{\,t}$ is the velocity vector of magnitude $v_0$ along direction $\theta_i$ and $\xi_i^t$ is a delta-correlated white noise representing perturbations, while $\eta$ is noise strength. Due to its simplicity this model lacks some realistic details but it was stimulating from the point of developing it further to obtain increasingly realistic simulational models.

As an extension of the above model Chaté and coworkers[82,83] added adhesive interactions in the form two-body repulsive-attractive forces and endowed the particles with size.

Thus the angle of direction of motion of particle $i$ is:

$$\theta_i^{t+1} = \arg\left[\alpha \sum_{j\sim i} \vec{v}_j^t + \beta \sum_{j\sim i} \vec{f}_{ij}\right] + \eta \xi_i^t, \quad \text{(Eq. 3)}$$

where $\alpha$ and $\beta$ control the relative weights of the two 'forces'. A hard-core repulsion at distance $r_c$ and an 'equilibrium' preferred
5 distance $r_e$ and $r_a$ attraction distance are ensured as:

$$\vec{f}_{ij} = \vec{e}_{ij} \begin{cases} -\infty & \text{if } r_{ij} < r_c, \\ \frac{1}{4}\frac{r_{ij}-r_e}{r_a-r_e} & \text{if } r_c < r_{ij} < r_a, \\ 1 & \text{if } r_a < r_{ij} < r_0 \end{cases} \quad \text{(Eq. 4)}$$

where $r_{ij}$ is the distance between particles $i$ and $j$, while $\vec{e}_{ij}$ is the unit vector along the segment going from $i$ to $j$.

The third basic model we describe here was proposed already to
10 describe how adhesive cells, having a finite size and a reorientation mechanism, move together.[2] In this model short-range attractive and repulsive intercellular forces are suggested to account for the organization of motile cells into coherent groups. Instead of applying an explicit averaging rule the model cells
15 (self-propelled particles, SSP) adjust their direction toward the direction of the net force acting on them [Eq. 4]. In this 2-dimensional flocking model, N SSPs move with a constant velocity $v_0$ in the direction of the unit vector $\vec{n}_i(t)$. In addition, independent of this active motion, cell pairs $i$ and $j$ also
20 experience an intercellular force $\vec{F}(\vec{r}_i, \vec{r}_j)$ which moves the cells' positions $\vec{r}_i(t)$ with a mobility $\mu$. Thus the motion of cell $i$ is described by the following equation:

$$\frac{d\vec{r}_i(t)}{dt} = v_0 \vec{n}(t) + \mu \sum_{j=1}^{N} \vec{F}(\vec{r}_i, \vec{r}_j) \quad \text{(Eq. 5)}$$

The direction of the unit vector $\vec{n}_i(t)$ is $\theta_i^n(t)$, which is
25 assumed to align with the physical total displacement $\vec{v}_i(t) = d\vec{r}_i(t)/dt$ with a relaxation time $t$, given by the following equation:

$$\frac{d\theta_i^n(t)}{dt} = \frac{1}{\tau}\delta(\vec{n}_i(t), \vec{v}_i(t)) + \xi \quad \text{(Eq. 6)}$$

In Eq. 6 the angle between the direction vector $\vec{n}_i(t)$ and velocity
30 vector $\vec{v}_i(t)$ is denoted by $\delta$ and imperfect alignment is represented by a noise term $\xi$.

Although the above models are formulated for the two-dimensional case (sheet migration) it is possible to extend them to
35 three-dimensional cases.

**Modeling of sheet migration**

A two-dimensional model of collective motion was developed for the sheet migration of keratocytes[2], detailed in the section 'Collective cell motion in vitro'.

40

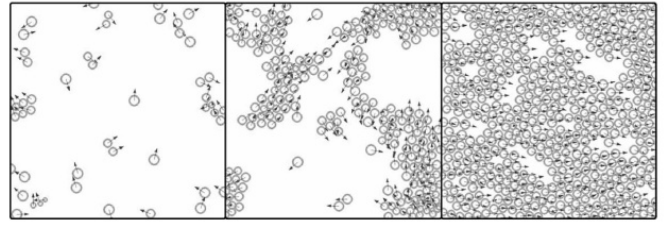

45 **Fig. 14** Computer simulations obtained by solving Eqs 5 and 6 for different particle densities. In agreement with the observations, the model exhibits a continuous phase transition from disordered to ordered phase. Also see Reference video 1. From Szabó et al., (2006) with permission of Phys Rev E.

50 The typical simulation results obtained by solving Eqs 5 and 6 with periodic boundary conditions, and shown in Figure 14, are in agreement with observations on sheet migration (**Fig. 1**), exhibiting a continuous (second order) phase transition from disordered to ordered phase as a function of increasing cell
55 density used as control parameter.

**Modeling of streaming in cell monolayers**

Streaming in monolayers was modeled by Czirok and coworkers[4,16] using Cellular Potts Model (CPM, also called
60 Glazier-Grainer-Hogweg Model), a widely used representation of individual cells and their adhesion. In the CPM approach, a goal function ('energy') is assigned to each configuration of cells. The goal function guides the cell behavior by distinguishing between favorable (low $u$) and unfavorable (high $u$) configurations as:

65
$$u = \sum_{\langle x,x'\rangle} J_{\sigma(x),\sigma(x')} + \lambda \sum_{i=1}^{N} \delta A_i^2. \quad \text{(Eq. 7)}$$

The first term in Eq. 7 enumerates cell boundary lengths. The summation goes over adjacent lattice sites. For a homogeneous cell population, the $J_{i,j}$ interaction matrix ($0 \leq i, j \leq N$) is given as:

$$J_{i,j} = \begin{cases} 0 & \text{for } i = j \\ \alpha & \text{for } i j > 0 \text{ and } i \neq j \text{ (intercellular boundary)} \\ \beta & \text{for } ij = 0 \text{ and } i \neq j \text{ (free cell boundary)} \end{cases}$$

70
$$\quad \text{(Eq. 8)}$$

The surface energy-like parameters $\alpha$ and $\beta$ characterize both intercellular adhesiveness and cell surface fluctuations in the model. The magnitude of these values determines the roughness of cell boundaries: small magnitudes allow dynamic, long and
75 hence curvy boundaries, while large magnitudes restrict boundaries to straight lines and thus freeze the dynamics.

*Cell polarity rule*

Cell polarity vector $\rho_k$ is assigned to each cell $k$. Then the
80 probability of elementary conversion steps advancing the cell center parallel to $\rho_k$ is increased as:

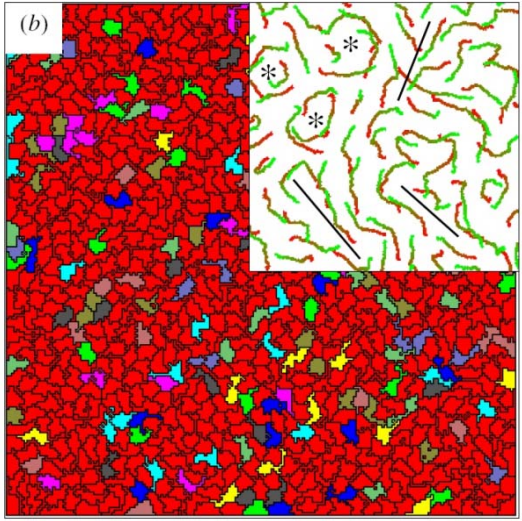

**Fig. 15** Simulation results with low cell adhesion and strong self-propulsion. The inset demonstrates cell trajectories, black lines separate cell streams moving in opposite direction, asterisks show vortices. Also see Reference videos 10 and 11. From Szabó et al. (2010) with permission of Phys. Biol.

$$\omega(a \to b) = P \sum_{k=\sigma(a),\sigma(b)} \frac{\Delta X_k(a \to b) p_k}{|p_k|}. \quad \text{(Eq. 9)}$$

where $P$ is a parameter setting the magnitude of bias and $\Delta X_k$ is the displacement of the center of cell $k$ during the elementary step considered.

*Polarity memory rule*

Cell polarity is updated by considering a spontaneous decay in polarity and a reinforcement from past displacements as:

$$\Delta p_k = -r p_k + \Delta X_k \quad \text{(Eq. 10)}$$

where $r$ is the rate of spontaneous decay and $\Delta X_k$ is the displacement of the center of cell $k$ during the Monte-Carlo steps (MCS) considered.

The cell polarity rule [Eq. 9] and the memory rule [Eq. 10] together constitute a positive feedback loop. The simulations have been performed applying periodic boundary conditions.

Results from simulations fit well with experimentally observed streaming patterns in endothelial monolayers: streaming motion, shear lines and vortices are seen, as shown in **Fig. 15** (also see Reference videos 10 and 11).

**Modeling of the role of leadership**

Based on experimental data from wound-healing assays with MDCK cell layers and measurable parameters of cell motion Lee et al.[84] developed a mathematical model incorporating the bulk features of single migrating cells and cell-cell adhesions.

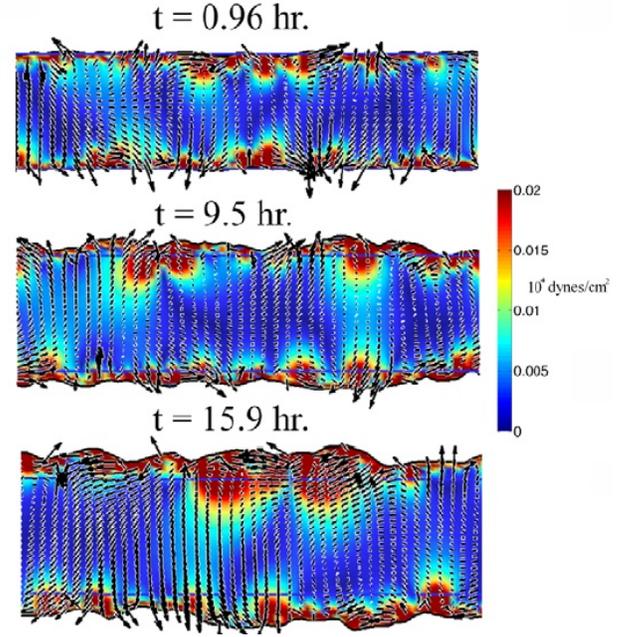

**Fig. 16** Complex flows and border progression in simulated two-dimensional wound healing assays. A characteristic time course from a simulation with an initial width of 200 μm showing the local velocity of the cells (black arrows) and the traction force exerted against the substrate (colormap). Inside the cell-filled region, the cells move with complex dynamics, which includes vortices and long-range correlations in the velocity field. The border advance is non-uniform and shows characteristics of a fingering-type instability. From Lee et al., 2011 with permission of PLOS Comput Biol.

The principal driving force in their model comes from the polarization of crawling cells: single crawling cells exert a dipole-distributed force distribution on the substrate. At the edge of the wound this force distribution acts like a pressure pulling the cells out into the cell-free region. Within the cell-filled region the force distribution causes instabilities leading to the experimentally observed flow fields including vortices, jets and fingering-like appearance of the moving boundary (**Fig. 16**).

In this model the cells are equivalent without differentiation into leaders and followers and as a result the boundary fingering is not as pronounced as what is observed experimentally. Cell-cell adhesions cause the monolayer to act like a viscoelastic fluid that is rigid on short timescales and flows on longer timescales.

This model's behavior such as the dynamics of the boundary advance matches well the data from experiments by Poujade et al (**Fig. 3**).[20] In various model simulations they have shown that wound healing may not require substantial biochemical signaling but the process may result only from the typical dynamics of motile cells while intercellular signaling only modifies the force production in cells at different distances from the boundary.

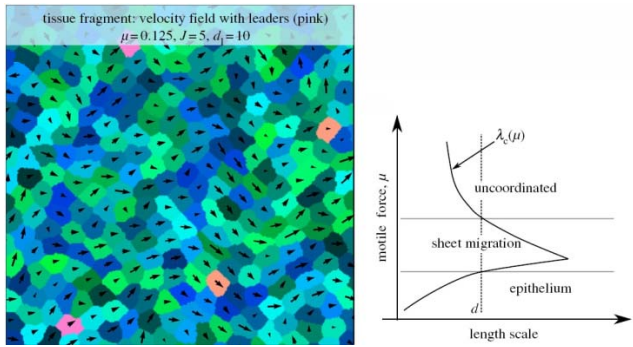

**Fig. 17** Left: An example of a tissue with a few leader cells (with pink/orange tone) whose polarity is constant and directed towards the right. Right: A sketch of the curve $\lambda_c(\mu)$ and its qualitative relationship with the different regimes of migration. For a given length scale $d$ associated with a constraint (distance between leaders, distance between boundaries), three regimes can be defined as $\mu$ increases: epithelium, sheet migration or uncoordinated. From Kabla (2011) with permission of J. R. Soc. Interface.

Using computer simulations Kabla[85] studied collective migration and its dependence on the number, motile force and cohesion energy of constituent cells. In these simulations, the degree of global coordination is quantified as mean velocity across the whole population normalized by the mean cell speed $(\langle v \rangle / \langle |v| \rangle)$ corresponding to an order parameter taking values from 0 (no order) to 1 (full coordination or sheet migration). This order parameter depends on motile force ($\mu$), cohesion energy ($J$) and system size. Typical length scales, $\lambda_g(\mu, J)$ can be identified corresponding to the largest system size where global coordination can arise spontaneously. For small populations of 10-100 uncoordinated cells it is shown that increase of motile force, $\mu$, or decrease of cohesion energy, $J$, could trigger sheet migration without need for specific signaling cues (**Fig. 17**).

The impact of leader cells and the integration of external directionality cues are also discussed. It is assumed here that leader cells are not concentrated at boundaries but scattered throughout the cell population. The susceptibility of the cell population to steering by 'informed' leader cells whose directional preference is based on e.g. sensing external cues depends on the distance between leader cells, $d_l$, (also manifested as leader cell density) and the collective effects in the bulk of the population.

Small relative number of leader cells (~1%) are sufficient to coordinate the whole cell population if $d_l < \lambda_l (\mu, J) \approx \lambda_c (\mu, J)$ where $\lambda_c$ is the correlation length of the average velocity field in the direction of local velocity, measured in the absence of leader cells.

As each leader cell influences the dynamics of the cells present within a domain of diameter $\lambda_c$ around it, global coordination can be achieved if the density of leader cells is larger than 1 for every domain $\lambda_c^2$. This way, large-scale coordination does not require explicit communication between leader and non-leader cells or long-range mechanical coupling through the substrate. Different regimes can be defined for a given correlation length scale as motile force is increased: (non-moving) epithelium, sheet migration and uncoordinated migration (**Fig. 17**).

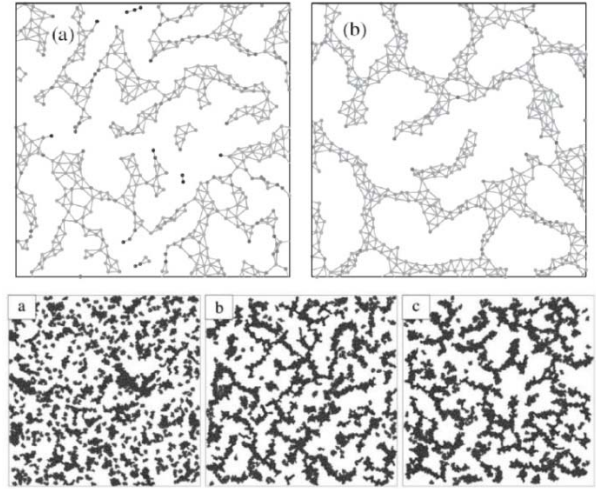

**Fig. 18** Computer simulations of early vasculogenesis by an agent-based model and a modified Cellular Potts Model.

Upper panel: Network formation in the agent-based model. Randomly placed $N = 500$ particles assemble into linear structures, detectable already within 30 minutes (a). At a sufficiently high particle density, a characteristic pattern size develops in five hours (b) with a combination of sprouting (branch extension) and coarsening (merger of adjacent branches). Connected dots represent Voronoi neighbor particles. Darkening gray levels indicate increasing local anisotropy. The simulation covered an area of $L = 0.7$ mm. From Szabo et al. (2007) with permission of Phys Rev Lett.

Lower panel: The Potts Model simulation reaches a stationary state where surface tension-driven coarsening is balanced by the formation of new sprouts. Configurations in the model are shown after 100 (a), 1000 (b), and 30,000 (c) Monte Carlo time-steps. As the structure factors averaged over 10 independent runs reveal, the emerged pattern does not change its statistical characteristics after 1000 steps (a). However, the resulting pattern is not frozen: branches still form and break up. Also see Reference video 12. From Szabo et al. (2008) with permission of Biophys J.

**Modeling of embryonic vascular network formation**

Early vascular network formation is a self-organizing process apparently lacking external prepatterns that vascular precursor cells could follow to get organized into a polygonal network, observed during in vivo development. Based on the simple assumption that endothelial cells preferentially attract to elongated cell structures, Czirok and coworkers performed computer simulations with both an agent-based model[86] and a modified Cellular Potts Model[43] and were able to create polygonal cell structures forming with a dynamism resembling the early vascular network of bird embryos (**Fig. 18,** also see Reference video 12). In the agent-based model, the simulated network of cells evolve into a quasistationary state in which the formation of new branches by preferential attraction mechanism is counterbalanced by coarsening of the network through merger of branches driven by surface tension. The characteristic size of the polygonal network depends on cell density.

An alternative mathematical model is based on the assumption that endodermal signaling exerting a paracrine effect on endothelial precursors is mediated by binding to the extracellular matrix deposited by the endothelial precursors.[87]

**Modeling of gastrulation in the zebrafish embryo**

Gastrulation of the zebrafish embryo was studied with the help of a numerical simulation by Arboleda-Estudillo et al.[5] In their simulation the migration of cells is mediated by 4 different force types: 1) a short-range repulsive, mid-range attractive spring force ($f_s$) representing cell adhesion; 2) a chemotactic force ($f_c$) modeling polarized migration; 3) a "Vicsek et. al. type" force, $f_v$, modeling collective migration as each cell attempts to align its direction with its neighbors; 4) noise force ($f_n$) modeling random migration.

For a system of $N$ cells, labeled by $r_i$, the system of $N$ coupled Langevin equations are numerically integrated:

$$b^{-1} (dr_i / dt) = \sum_{j \neq i} f_s + f_c + f_v + f_n. \quad \text{(Eq. 11)}$$

where $b$ is cell mobility. The simulations were performed with periodic boundary conditions in y direction. The results were consistent with experimental observations: mesendoderm cell groups with decreased cell-cell adhesion strength, and simulated cell groups with lower spring force both exhibited less directed and slower movement during collective migration. Cell-cell adhesion is hypothesized to decrease the variability of the movement path of individual cells during collective migration by coupling the cells and hence posing steric constraints.

**Modeling of collective migration of the posterior lateral line primordium of the zebrafish**

Based on experimental data from lateral line development in the zebrafish Streichan et al.[88] have devised a model integrating numerous known factors of the process (**Fig. 19**). They propose a dynamically established and maintained mechanism in which there is no need for an already established chemokine ligand gradient to direct the migration of a cell collective. In their model the cell collective actively modulates the isotropically expressed chemokine. The ligand is degraded and co-internalized with receptor, which reduces ligand concentration in the vicinity of the tissue. As the tissue moves it shapes the ligand distribution to an asymmetric profile resulting in a new mean gradient in ligand concentration in the direction of migration. Hence the collective's migration creates a length- and velocity-dependent polar gradient. Cells encode an initial symmetry breaking in their velocity to shape the chemokine ligand, initiate the traveling wave and maintain the preferred direction of motion.

The model makes predictions on the length-dependent dynamics of the lateral line primordium and the spatio-temporal dynamics of receptor-ligand interaction. Authors identify competition between the front and the rear arising from tissue extensions above a critical length and leading to deposition of cells as the collective migrates along.

**Modeling of neural crest cell migration and collective chemotaxis**

Collective chemotactic migration of groups of neural crest cells has been subjected to various modelling approaches. An agent-based model has been elaborated by Mayor and coworkers on the basis of the cellular and molecular mechanisms reported so far to underlie neural crest cell migration.

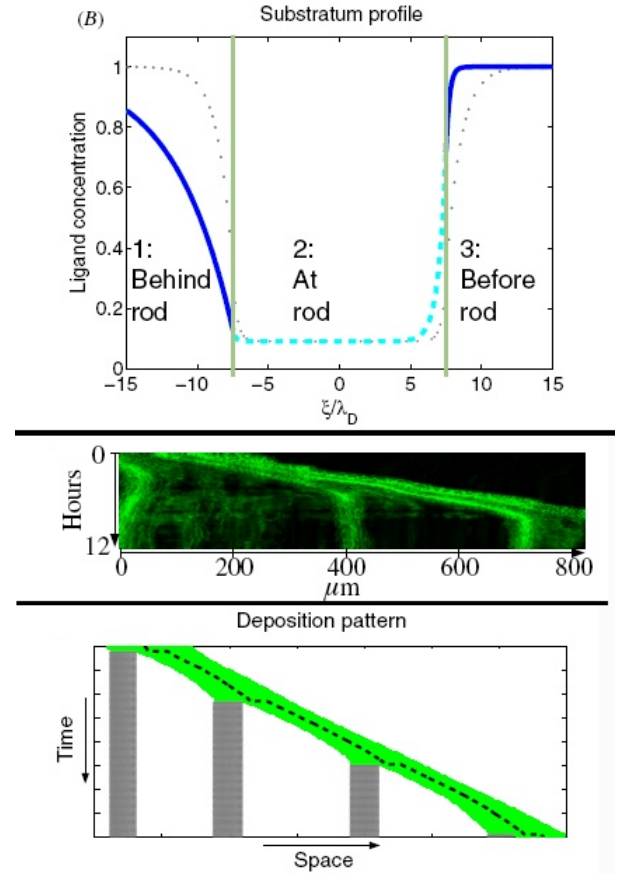

**Fig. 19** Upper panel: Typical ligand concentration in the vicinity of the rod with the $v > 0$ solution shown as dashed cyan line and in the free space shown as solid blue. The green lines denote the front and the rear of the rod. A strong gradient at the front of the rod is observed, whereas in the centre of the rod the new steady-state ligand concentration is reached. The dotted grey profile indicates the symmetric $v = 0$ solution.

Middle panel: Kymograph shows the temporal evolution of the fluorescence signal along a section through the maximum intensity projection of the lateral line primordium. Time is along the *y*-axis and the section's extension is along the *x*-axis. At 0 *μ*m, a neuromast deposition is shown: the fluorescence signal of deposited cells becomes stationary, i.e. parallel to the time axis, which corresponds to static cell groups. The front of the tissue continues migration as indicated by straight lines that form an obtuse angle with the *x*-axis. At about 400 and 700 *μ*m further cell depositions are observed.

Lower panel: Simulation of the elastic rod with deposition. Deposited parts are dotted grey, the rod is shown as solid green lines and the centre of mass of the rod as a dashed black line. The rod moves to the right and grows at a rate *η* until a critical length is reached, which leads to the deposition of cells. The remainder continues migration. The speed of the centre of mass decreases until a next deposition is observed. From Streichan et al. (2011) with permission of Phys Biol.

Importantly, this model does not assume neural crest cells to functionally differentiate into leaders or followers. The i) short-range repulsive interactions corresponding to contact inhibition of locomotion and ii) longer-range mutual attractive interactions among cells and iii) migration biased towards a chemotactic gradient have been implemented in the model.

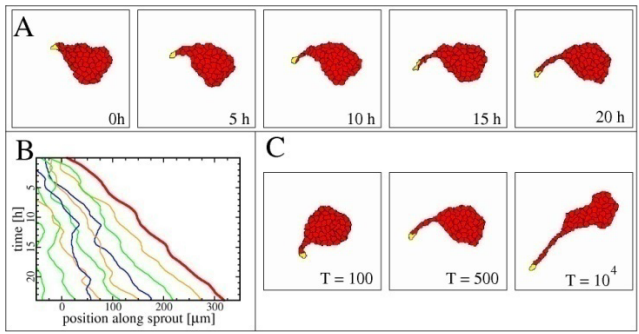

**Fig. 20** Computational model of multicellular sprout elongation: leader cell-initiated sprouting behavior in a computational model system with preferential attraction to elongated cells. A: typical time-course of sprout growth: the leader is slightly elongated, thus it pulls passive cells from the initial aggregate. The passive cells become elongated as well and attract further cells into the growing sprout. With sufficient supply of cells, the expansion can continue for an extended time period. B: cell trajectories along the sprout direction reveal cells entering the sprout as well as changes in cell order due to differential motion in the sprout. C: persistence time of polarity defines sprout shape and length, through the polarity persistence parameter $T$. When the leader cell is more persistent, longer and straighter sprouts form. From Szabo et al., (2010) with permission of Math Mod Nat Phenom

Corresponding simulations have shown that these three are sufficient to reproduce the group migration dynamics of NC cells observed experimentally.[32] An alternative agent-based model of the chain migration of neural crest cells is based on the assumption that leaders and followers differentiate from a homogeneous population NC cells. Leaders are directionally biased towards a target and followers move towards the least resistance in the extracellular matrix opened up by leaders while contact guidance by fillopodial interactions among cells further helps them follow the leaders.[89]

**Modeling of vasculogenesis by biophysical mechanism**

The basic process of vascular network formation is the initiation and development of multicellular sprouts maturated into blood vessels later on. Szabo et al. studied sprouting in a simplified in vitro system without chemokines.[43] Motivated by experimental findings they have developed a model based on the assumption of preferential adhesion to elongated cells (**Fig. 20**).[90] In their modified Cellular Potts Model cells prefer to be adjacent to other stalk cells rather than staying in the aggregate (see Reference video 13). The presence of persistently moving tip cells and the preferential adhesion assumption are together sufficient to generate expanding sprouts in computer simulations with this model (reviewed in:[91,92,16]).

Another approach on modeling angiogenic network formation based on purely local mechanisms was elaborated by Deutsch and coworkers.[93]

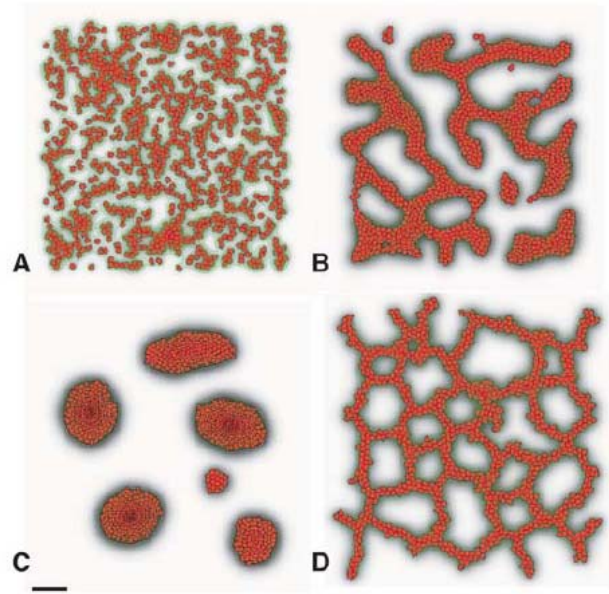

**Fig. 21** Endothelial cell aggregation; simulation initiated with 1000 scattered cells. (A) After 10 Monte Carlo steps (MCS) (~5 min). (B) After 1000 MCS (~8 h). (C) After 10,000 MCS (~80 h). (D) Contact-inhibited chemotaxis drives formation of vascular networks. Scale bar: 50 lattice sites ($\approx$ 100 µm). Contour levels (green) indicate ten chemoattractant levels relative to the maximum concentration in the simulation. Grey shading indicates absolute concentration on a saturating scale. From Merks et al. (2008) with permission of PLoS Comput Biol.

In their lattice-gas cellular automaton model the increased movement coordination and cell-cell adhesion of simulated cells in response to homogeneous growth factor (VEGF) stimulation is sufficient to result in angiogenic sprouts resembling the image data from in vitro experiments with endothelial cells.[94] In particular, this model does not assume changes in contact guidance or extracellular matrix remodeling or spatial gradient of growth factor.

**Modeling of vasculogenesis by chemotaxis**

Vascular sprouting can be approached as a process in which cells secrete a diffusible chemoattractant morphogen thereby inducing autocrine chemotactic signaling.[44,45,46] Glazier and coworkers investigated this mechanism using a computer model.[95,96]
In their Cellular Potts Model they assume finite compressibility of cells and as a result effective pressure is developed within the aggregate formed by cells migrating toward the chemoattractant produced by the cells while the steepest gradient is at the surface. Chemotaxis and pseudopod formation by a cell is assumed to be inhibited by surrounding cells through a mechanism called 'contact inhibition'. If random motility fluctuations move a cell away from the cluster it will sense a weaker chemoattractant gradient and the pressure of the compressed cells continues to push the same cell outward, while pseudopod formation of the cell is released from contact inhibition.

Simulations with this model yield sprouts and network formation (**Fig. 21**) and show that the sprouting process is facilitated by cells' finite cell size, the presence of elongated cells and increased chemotactic sensitivity.

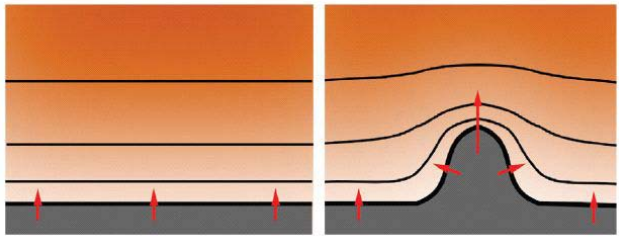

**Fig. 22** Mullins–Sekerka instability develops when the dynamics of a diffusive field is fast and a stronger gradient accelerates the movement of the interface. In such systems the tip of a 'sprout' senses larger gradients in the 'updated' concentration field, i.e. in the field that is adapted to the altered shape of the interface. Hence the sprout elongates as long as it can effectively reduce the concentration of the chemoattractant at the tip. Concentration is indicated by orange color, and selected concentrations by black contour lines while red arrows with proportionate lengths point up the gradient that a cell senses. From Czirok (2013) with permission of Wiley Interdiscip Rev Syst Biol Med).

The model also assumes that the main source of the chemoattractant ligand is the endothelial cells themselves. This assumption, however, conflicts with experimental data on the production and abundance of VEGF, the candidate chemoattractant morphogen. This contradiction can be overcome by assuming a secondary mechanism creating a gradient from even distribution of VEGF by sequestration.

A recent computational study[97] has demonstrated that if production of soluble VEGFR1 is proportional to endothelial cell density while VEGF production is uniform and high, a gradient of VEGF-induced signaling through VEGFR2 receptor is established along the sprout surface with highest signaling activity at the sprout tip. Experimental data on a secreted diffusible VEGF receptor support the existence of such a mechanism.[47,48]

The patterning process based on extension of a structure up the gradient of an external diffusible factor has an established theory. If the concentration of the diffusible factor is kept low at the interface of the cell aggregate while it is uniform high far from it, and if concentration is proportional to the local curvature of the interface, such setting results in classic Mullins-Sekerka instability, shown to be responsible for the formation of dense branching patterns in various physical systems. The Mullins-Sekerka instability makes the smooth surface unstable: a spontaneous outgrowth with higher curvature will sense a steeper gradient, which accelerates its growth, provided that adaptation of the gradient is slower than such growth (**Fig. 22**). This way the instability triggers a spontaneous tip-splitting process creating structures with characteristic branching morphology.

The branching process is balanced by the fact that very thin sprout with very large curvature at its tip cannot reduce the diffusible factor concentration so efficiently due to its small size and thus the gradient will become shallower, resulting in slower growth. Eventually, optimal branch width can develop with thicker branches splitting and thinner branches slowing down and growing laterally. While experimental verification is yet to be established, the patterning mechanism based on diffusible secreted inhibitor is a promising approach to understand vascular sprouting.

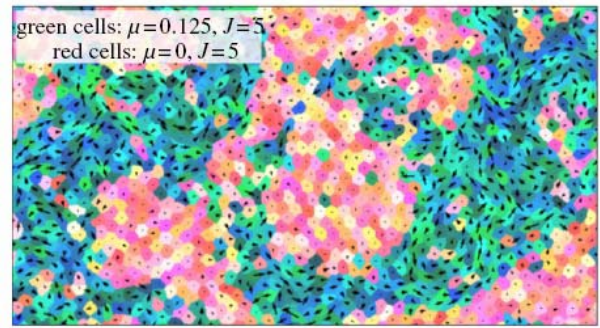

**Fig. 23** Simulated segregation of motile and non-motile cells. A snapshot of the simulated segregating tissue of motile and non-motile cells at $t = 10^6$ MCS (Monte-Carlo steps). Membrane tension, $J$, and motile force, $\mu$, of cells are indicated. From Kabla (2012)[25] with permission of J R Soc Interface.

### Modeling of cellular segregation

Several computational models exist that attempt to explain and reproduce the experimentally observed segregation processes in various systems. A widely accepted model based on the Potts model and the idea of differential cell adhesion was developed by J Glazier and co-workers, later termed as Glazier-Graner-Hogeweg model or Cellular Potts Model.[68] Variants of this model have been successfully employed in simulation works up to the present days.

### Impact of motility on segregation

Dynamic segregation in 2-dimensions was studied by Kabla using Cellular Potts Model simulations with self-propelled motile and non-motile cells characterized by identical adhesive properties.[84] Segregation efficiency has been found to depend on the motile forces controlling cell speed, and efficiency reaches maximum at motile forces close to the threshold required for streaming transition. It is also shown by these simulations that differences in motility are sufficient to drive the segregation of cell populations even without difference in adhesion and as a result motile cells will surround the islands of non-motile cells (**Fig. 23**).

Recently, Nakajima and Ishihara used Cellular Potts Model simulations to study the dynamics of the segregation of mixtures of non-self-propelled cell types with diffusive motion.[51] They have found that the increase in the size of segregated domains follows power law and the growth exponent is n = 1/3 for mixtures with 1:1 initial ratio of cell types where segregation proceeds via smoothing of the domain boundary. This is in contrast with previous works with CPM on smaller simulated systems displaying slower logarithmic growth for domain size.[69,98] CPM simulations with self-propelled cell types characterized by identical adhesive interactions as for the simulations by Nakajima and Ishihara[51] also yield domain growth exponent n = 1/3 (A. Czirók, personal communication).

**19**

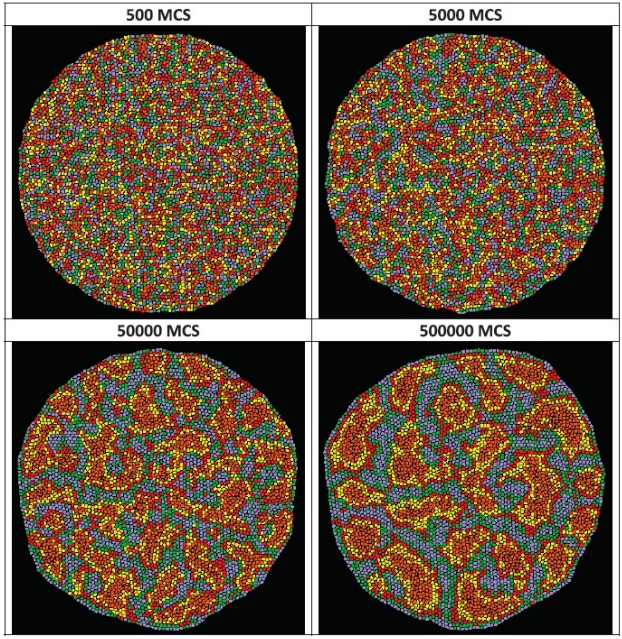

**Fig. 24** Clustering dynamics of cells with different adhesion characteristics. Snapshots taken from a 5000 cell aggregate simulation with five levels of cadherins showing the dynamics of cluster formation. Time points are denominated as Monte-Carlo steps (MCS). From Zhang et al. (2011) with permission of PLoS One.

Using Brownian dynamics simulations McCandlish et al. studied dense mixtures of self-propelled and passive rod-like particles in 2-dimensions where only excluded volume interactions can occur.[99] Adhesion properties do not play a role here, particles only differ in motility. Spontaneous segregation of the two particle species generates a rich array of dynamical domain structures with properties depending on particle shape and propulsion velocity or the combination of these two in the form of particles' Péclet number, a measure similar to the directional persistence of live cells.

**Impact of adhesion on segregation**

The role of adhesion in cell segregation was studied by Zhang et al. using Cellular Potts Model for simulations.[100] In their model they consider variations in the distribution of adhesion molecules per cells. The speed of segregation is found to increase strongly with interfacial tension that depends on the maximum difference in the number of cadherin adhesion molecules per cell and the reaction-kinetic models of cadherin binding (**Fig. 24**).

Qualitative description of the dynamical features and the geometry of cell segregation depending on intercellular adhesion parameters was provided by Voss-Böhme and Deutsch using a stochastic interacting particle model.[101] In this model the hierarchy of segregation is determined by the strengths of adhesive interactions between cells and the boundary.

In a unique paper combining experimental data and modeling Krieg et al.[58] studied the role of cell-cortex tension and adhesion in the segregation of germline progenitors of the zebrafish. Carrying out simulations using Cellular Potts Model with cell adhesion and cell-cortex tension data derived from experiments they could reproduce the experimentally observed final configurations of segregating germ line progenitor cell types.

**Segregation by collective motion and adhesion**

To study cell sorting events Chaté and coworkers developed a model[63] combining the collective motion model of Vicsek et al.[81,1] with the differential adhesion hypothesis (DAH). In their model $N$ particles move in 2-dimensional space with constant velocity $v_0$. The velocity and the angle of orientation of particle $n$ at time $t$ is denoted by $\vec{v}^t_n$ and $\theta^t_n$, respectively. The new orientation $\theta^{t+1}_n$ of particle $n$ is:

$$\theta^{t+1}_n = \arg\left[\sum_m \left(\alpha_{nm}\frac{\vec{v}^t_m}{v_0} + \beta_{nm} f^t_{nm} \vec{e}^t_{nm}\right)\right] + \vec{u}^t_n \quad \text{(Eq. 12)}$$

where $f^t_{nm}\vec{e}^t_{nm}$ is the force exerted by particle m on particle n along the direction $\vec{e}^t_{nm}$ pointing from particle m to n.
Noise is taken into account by $\vec{u}^t_n$ is a unit vector with random, uniformly distributed orientation.

Here, $\alpha_{nm}$ and $\beta_{nm}$ are control parameters: $\alpha$ controls the relative weight of the alignment interaction and $\beta$ shows the strength of the radial two-body forces $f_{nm}$, defined as:

$$f_{nm} = \begin{cases} \infty & \text{if} \quad r_{nm} < r_c \\ 1 - \dfrac{r_{nm}}{r_e} & \text{if} \quad r_c < r_{nm} < r_0 \\ 0 & \text{if} \quad r_{nm} > r_0 \end{cases} \quad \text{(Eq. 13)}$$

that is for distances smaller than a core radius $r_c$ it is a strong repulsive force, around the equilibrium radius $r_e$ it is a harmonic-like interaction, whereas for distances larger than the interaction range $r_0$ it is set to zero.

Having the classic experiments with Hydra cells in mind, authors defined two kinds of particles, "endodermic" and "ectodermic", denoted by 1 and 2, respectively. Accordingly, $\beta_{11}$ and $\beta_{22}$ stand for adhesion within the given cell type, whereas $\beta_{12} = \beta_{21}$ account for symmetric inter-cell-type interactions. Differential adhesion is described by different beta values for symmetric interactions between different cell types. The simulations were performed with cells on a square domain with linear size several magnitudes larger than cell size. Figure 25 shows snapshots from the evolution of the segregation process. Simulation results are in agreement with experiments of Rieu et al. with dissociated ectodermal and endodermal cells of *Hydra viridissima*.[102]

In this model, segregation is characterized by an index, $\gamma$, showing the average ratio of dissimilar cells around a cell, for either cell types. This index is decreasing as segregation proceeds and it is expected to approach zero in large systems.

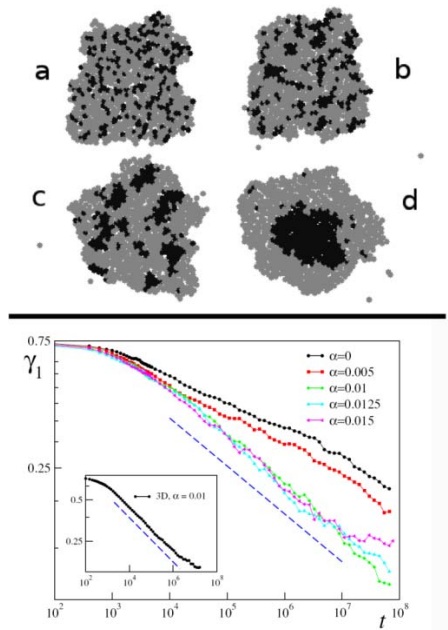

**Fig. 25** Segregation dynamics of simulated ectoderm and endoderm cells.

Upper panel: Cell sorting of 800 cells. The endodermal and ectodermal cells are represented by black and gray circles, respectively. (a) The initial cluster with mixed cell types. (b) The cluster after 3000 time steps and (c) is taken at $t = 3 \times 10^5$. Clusters of endodermal cells form and grow as time passes by. (d) At $t = 2 \times 10^6$ a single endodermal cluster is formed, but isolated cells remain within the ectoderm tissue, in agreement with experiments of Rieu et al., 1998.[102]

Lower panel: Cell sorting in two dimensions from a random, roughly circular initial aggregate of N = 6400 cells in a proportion of 1:3 endodermic to ectodermic cells. Evolution of the segregation index, γ, for different α values. The dashed line has a slope $-\lambda = -0.18$. Inset: Same in three dimensions but with $\alpha = 0.01$ and $\beta_{11} = 8.3$. The dashed line has a slope -0.16. From Belmonte et al. (2008)[63] with permission of Phys Rev Lett.

Authors have found that segregation is characterized by algebraic scaling laws and introducing even a moderate amount of local coherent motion will considerably speed up the segregation process (**Fig. 25**).

A variant of this computational model has been published by Beatrici and Brunnet investigating the segregation of self-propelled particles in 2 dimensions, driven by differences only in motility but not in adhesion.[103] In this model, the faster cells envelope the slower cells forming islands as segregation proceeds.

Further developing the model collective motion of Vicsek and coworkers[2] Mones et al. have recently carried out simulations of the segregation behavior of 'self-propelled' particle types compared with that of 'noise-driven' particle types.[62] To represent interactions with neighbors, particle types were assigned characteristically different two-body attraction/repulsion forces based on experimental data with live cells. Noise-driven particle types, endowed with inherent random motion and no ability to have information from neighbors, segregate with similar dynamism as particles in Potts Model simulations by Nakajima and Ishihara,[52] i.e. exponent values ~1/3 characterize the growth of segregated domains. As a contrast, self-propelled particles with persistent motion and the ability to align their motion to neighbors in response to impact by neighbors segregate much faster, with growth exponents ~1, and their dynamism resemble to earlier observations of two-dimensional and three-dimensional segregation of cells in culture.[51,56]

Although three-dimensional simulations and models have been deployed in other fields of cell motion,[79] approaching the phenomenon of three-dimensional segregation with such models remains an area yet to be explored by computational modelers.

## Conclusions

The way we approach and understand the events of developmental biology such as collective cell motion and pattern formation by multicellular segregation is gradually shifting from a descriptive view towards a causative understanding of the mechanisms. To facilitate this understanding, integrative biological attempts have been successfully employing various approaches ranging from experimental embryology to statistical physics. The introduction of computational models simulating the behavior of complex developmental systems can also effectively facilitate the way we interpret them. Combination of multi-disciplinary approaches with experimental data can help us design more focused experimental tests or predict yet unseen outcomes. This way they can even further extend our understanding of the dynamic organization of multicellular biological systems.

## Acknowledgement

We acknowledge support from the EU FP7 ERC COLLMOT GRANT No: 227878. We thank A. Czirok (Eötvös University, Dept. Biological Physics) for useful consultations.

## Notes and references

[a] Department of Biological Physics, Eötvös University, Budapest, Hungary. Fax: 36 13722757; Tel: 36 13722795; E-mail: emehes@angel.elte.hu
[b] MTA-ELTE Statistical and Biological Physics Research Group of the Hungarian Academy of Sciences, Budapest, Hungary. Fax: 36 13722757; Tel: 36 13722795; E-mail: vicsek@hal.elte.hu

† Electronic Supplementary Information (ESI) available:

## Video references

Reference video 1

Time-lapse sequences of phase contrast images showing the motility of fish epidermal keratocyte cells at three different densities. Each video is 4 hours long. Robust collective behavior can be observed as the density of cells reaches a critical value around $5 \times 10^{-4}$ cell per square microns. This spectacular ordering phenomenon resembles the well-known flocking of fish or birds.